\documentclass[pra,twocolumn,showpacs,superscriptaddress]{revtex4-1}
\usepackage{ulem}
\usepackage{graphicx}
\usepackage{amsmath}
\usepackage{amssymb}
\usepackage{latexsym}
\usepackage{fancyhdr}
\usepackage{array}
\usepackage{amsfonts}
\usepackage{mathrsfs}
\usepackage{mathtools}
\usepackage{color}
\usepackage{verbatim}
\usepackage{physics}
\usepackage[caption=false]{subfig}
\usepackage{natbib}
\usepackage{enumitem}
\usepackage{bbold}
\usepackage[colorlinks=true,urlcolor=blue,citecolor=blue,linkcolor=blue]{hyperref}

\usepackage{float}

\begin{document}
\title{General mapping of multi-qu$d$it entanglement 
conditions to non-separability indicators for quantum optical fields}

\author{Junghee Ryu}
\affiliation{Centre for Quantum Technologies, National University of Singapore, 3 Science Drive 2, 117543 Singapore, Singapore}

\author{Bianka Woloncewicz}
\affiliation{International Centre for Theory of Quantum Technologies (ICTQT), University of Gdansk, 80-308 Gdansk, Poland}

\author{Marcin Marciniak}
\affiliation{Institute of Theoretical Physics and Astrophysics, Faculty of Mathematics, Physics and Informatics, University of Gda\'{n}sk, 80-308 Gda\'{n}sk, Poland}

\author{Marcin Wie\'{s}niak} 
\affiliation{Institute of Theoretical Physics and Astrophysics, Faculty of Mathematics, Physics and Informatics, University of Gda\'{n}sk, 80-308 Gda\'{n}sk, Poland}
\affiliation{International Centre for Theory of Quantum Technologies (ICTQT),
University of Gdansk, 80-308 Gdansk, Poland}

\author{Marek \.{Z}ukowski}
\affiliation{International Centre for Theory of Quantum Technologies (ICTQT),
University of Gdansk, 80-308 Gdansk, Poland}

\date{\today}

\begin{abstract}
We show that any multi-qudit entanglement witness leads to a non-separability indicator for quantum optical fields, which involves intensity correlations. We get, e.g., {\it necessary and sufficient} conditions for intensity or intensity-rate correlations to reveal polarization entanglement. We also derive separability conditions for experiments involving multiport interferometers, now feasible with integrated optics. We show advantages of using intensity rates rather than intensities, e.g., a mapping of Bell inequalities to ones for optical fields.  The results have implication for studies of non-classicality of ``macroscopic" systems of undefined or uncontrollable number of ``particles".
\end{abstract}

\maketitle

\newcommand{\avg}[1]{\langle #1\rangle}
\newcommand{\cre}[1]{a^{\dagger}_{#1}}
\newcommand{\ani}[1]{a_{#1}}
\newcommand{\inner}[2]{\langle#1 | #2 \rangle}
%\newcommand{\ket}[1]{ | #1 \rangle}
%\newcommand{\bra}[1]{ \langle #1 | }

%--------------------------------------------------------------------------------
%--------------------------------------------------------------------------------
%\section{Introduction}
Non-classicality 
due to entanglement initially was studied using quantum optical  
multiphoton interferometry, see e.g.,~\cite{PAN}. 
The experiments were constrained  to defined photon number 
states, e.g., the two-photon polarization singlet~\cite{ASPECT}. 
This includes Greenberger-Horne-Zeilinger (GHZ)~\cite{GHZ} inspired multiphoton interference,
 with an interpretation that each detection event signals one photon.
Spurious events of higher photon number counts contributed  to a lower interferometric 
contrast.
Still, states of undefined photon numbers, e.g., the 
squeezed vacuum, can be entangled~\cite{BANASZEK, BOUW, MASZA-REVIEW}.

This form of entanglement of quantum optical fields served e.g., to show that a strongly pumped two-mode (bright) squeezed state allows one 
to directly refute the ideas of EPR \cite{EPR}, as it approximates their state, and a form of Bell's Theorem can be shown for it \cite{BANASZEK}. The trick was to use displaced parity observables.
Recently it has been shown that this is also possible for four-mode bright squeezed vacuum \cite{ROSOLEK}, which can be produced via type II parametric down-conversion, see e.g  \cite{BOUW, MASZA-REVIEW}. In this case the state approximates a tensor product of two EPR states, and interestingly can also be thought of as a polarization ``super-singlet" of undefined photon numbers \cite{DURKIN}. The approach of Ref. \cite{ROSOLEK}  used (effectively) intensity observables, which are less experimentally cumbersome.

With the birth of quantum information science and technology, entanglement became a resource. We have an extended literature on detection of entanglement for systems of finite dimensions, essentially ``particles", see e.g., \cite{HORODECKIS}.  It is well known that not all entangled states violate Bell inequalities. 
Still there is theory of entanglement indicators, called usually witnesses, which allow to detect entanglement, even if a given state for finite dimensional systems (essentially, qu$d$its) does not violate any known Bell inequalities. The case of two-mode entanglement for optical fields was studied in trailblazing papers of 
\cite{SIMON, DUAN}, which discussed ``two-party continuous variable systems",  and with a direct quantum optical formalism in \cite{HILLERY}.  The entanglement conditions reached in the papers did not involve intensity correlations.

An entanglement condition for four-mode fields, which was borrowing ideas from two spin-1/2 (two-qubit) correlations,  involved correlations Stokes operators and  was first discussed in \cite{BOUW}.  The resulting indicator was used to measure  efficiency of an ``entanglement laser". 
The output of the ``laser" was bright four-mode vacuum. 
We shall present here the most extensive generalization of such an approach, i.e., entanglement indicators for optical fields which are derivatives of multi-qudit entanglement witnesses involving intensity correlations.
In Supplementary Material~\cite{supp} we give examples of entanglement conditions based on such an approach. Some of them are more tight versions of the entanglement conditions  mentioned above. 

As a growing part of the experimental effort is  now directed at non-classical 
features of bright (intensive, ``macroscopic") beams of light, e.g., \cite{Lamas01, BOUW-2, Eckstein11, Iskhakov12, Iskhakov13, Kanseri13, Spasibko17} so the time is ripe for a comprehensive study of such entanglement conditions. All that may lead to some new schemes in quantum communication and quantum cryptography, perhaps on the lines of Ref. \cite{DURKIN}.
The emergence of integrated optics allows now to construct stable multiport interferometers \cite{Mattle95, Weihs96, Peruzzo11, Meany12, Metcalf13, Spagnolo13, Carolan15, Schaeff15}, and is our motivation of going beyond two times mode case.
% {\color{blue} However, compare to our knowledge of entanglement indicators for systems of finite dimensions, we have less experiences for optical fields, %especially macroscopic lights of undefined number of photons with respect to efficient measurements, noise robustness optimal conditions, and so on. It %results in a difficulty to derive an indicator for such entanglement.}

We present a theory of 
mapping multi-qu$d$it entanglement witnesses~\cite{HORODECKIS}
into entanglement indicators for quantum optical fields,
which employ intensity 
correlations or correlations of intensity rates. By intensity rates we mean the ratio of  intensity at a given local detector and the sum of intensities at all  local detectors (in some case the second approach leads to better entanglement detection). The method may find applications also  in studies of non-classicality of correlations in ``macroscopic" many-body quantum systems of undefined or uncontrollable
number of constituents, e.g., Bose-Einstein condensates \cite{SORENSEN},  other specific states of cold atoms \cite{POLZIK,sorensenPRL}.

The essential ideas are presented 
for polarization measurements by two observers  and the most simple model of 
intensity observable: photon-number in the observed mode.
 Next, we present further generalization of our approach, and  examples employing
specific indicators involving intensity correlations 
%of intensities 
for unbiased multiport interferometers. 
We discuss generalizations 
to multi-party
entanglement 
indicators.  
We show that the use of rates leads to   a modification of quantum optical 
Glauber correlation functions, which gives a new
tool for studying non-classicality,  and that it also  gives a   
  general method of mapping 
standard Bell inequalities into ones 
for optical fields.

\begin{figure}[t]
\centering
\includegraphics[width=8cm]{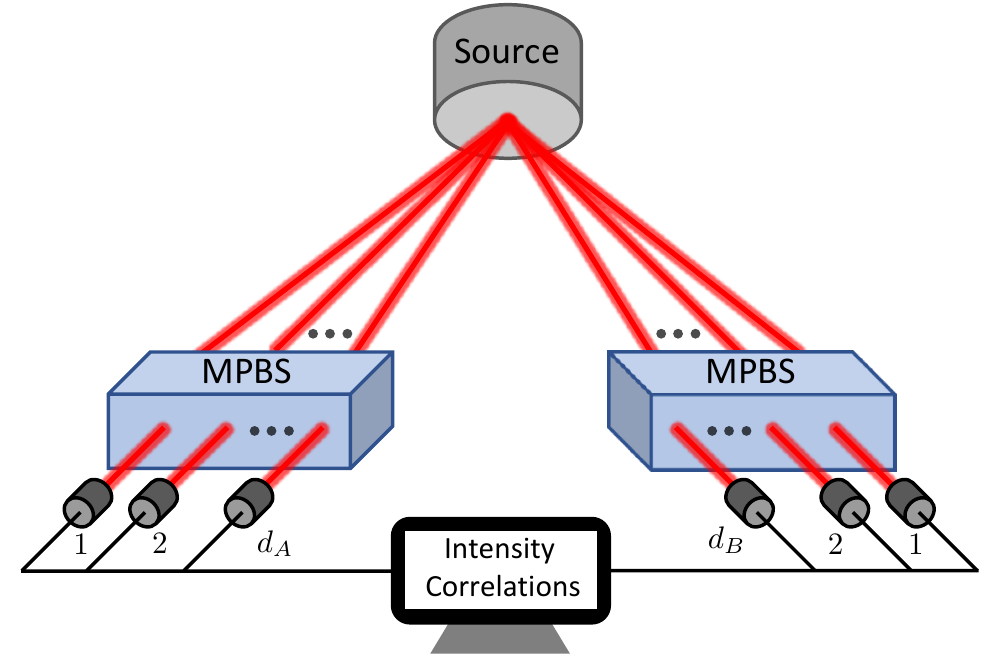}
\caption{The experiments (two parties). Two multi-mode beams propagate to two spatially separated measurement stations.
Each station consists of a $d$ input $d$ output tunable  multi-port beamsplitter-interferometer (MPBS) and detectors at its outputs.
  For polarization measurements put $d_A=d_B=2$, and treat the paths as polarization modes.}
\label{fig:exp}
\end{figure}

%{\it Generalities.}---
%Entanglement is most interesting 
%when it is observed  by spatially separated parties. 
We discuss  spatially 
separated stations, $X=A,B,...$  
%In detail we shall
%discuss only two spatially separated stations (parties), $X=A,B$
with
(passive) interferometers  
 of $d_X$ input and output ports, FIG.~\ref{fig:exp}.
In each output  there is a detector which measures
 intensity. 
One can assume either a pulsed source, 
sources acting synchronously~\cite{ZUK-ZEIL-WEIN, KALTENBAEK} 
or that the  measurement is performed within a short time gate. 
Each time gate, or pulsed emission, 
is treated as a repetition of the experiment building up averages of observables. 

{\it Stokes parameters.}---For the 
description of polarization of light, 
the standard approach uses Stokes parameters.
Using  the photon numbers they read
$
\langle \hat \Theta_{j}\rangle = \langle \hat{a}_{j}^\dagger \hat a_{j} - \hat a_{{j}_{\perp}}^\dagger \hat a_{{j}_{\perp}}\rangle, 
$
where $j, j_{\perp}$ 
denote a pair of orthogonal  
polarizations  of one of three mutually unbiased 
polarization bases $j = 1,2,3 $, e.g., 
$\{ {H}, {V}\}, \{ {45^{\circ}}, {-45^{\circ}}\}, \{ {R}, {L}\}$. 
The zeroth parameter $
\langle \hat \Theta_0 \rangle $ is the  total intensity:
$\langle \hat N\rangle = \langle \hat a_{j}^\dagger \hat a_{j} + \hat a_{{j}_\perp}^\dagger \hat a_{{j}_\perp} \rangle.$
Alternative {\it  normalized} Stokes observables were  studied
by some of us~\cite{zuko, zuko1, zuko2}. They  were first intorduced in \cite{HE}, however a different  technical approach was used.
Following \cite{zuko} one can put 
$
\langle \hat S_{j} \rangle = \langle \hat \Pi\frac{(\hat a_{j}^\dagger \hat a_{j} - \hat a_{{j}_\perp}^\dagger \hat a_{{j}_\perp})}{\hat N} \hat \Pi\rangle,
$
 and  $\langle \hat S_0 \rangle  = \langle \hat \Pi\rangle$,
where $\hat \Pi = \mathbb{1} - \ketbra{\Omega}$ and $|\Omega\rangle$ 
is the vacuum 
state for the considered modes, 
$\hat{a}_j|\Omega\rangle = \hat{a}_{j_{\perp}}|\Omega\rangle =0$.
Operationally, in the $r$-th run of an experiment, we register photon numbers in the two exits of a polarization analyzer, $n^r_j$ and $n^r_{j_\perp}$, and divide their difference by their sum. If  $n^r_j + n^r_{j_\perp}=0$ , the value is put as zero. This does not require any additional measurements, only the data are differently processed than in the standard approach. In~\cite{zuko, zuko1, zuko2} examples of two-party entanglement conditions and Bell inequalities using normalized Stokes operators were given. Here we  present a  general approach.

%%%%%%%%%%%%%%%%%%%%%%%%%%%%%%%%%%%%%%%%

{\it Map from two-qubit entanglement witnesses 
to entanglement indicators for fields involving Stokes parameters.}---Pauli operators 
$ \vec{\sigma}
= (\hat \sigma_1, \hat \sigma_2, \hat \sigma_3)$ and 
$\hat \sigma_0 = \mathbb{1}$ 
form a basis in the real space of one-qubit observables. 
Thus,  any two-qubit entanglement 
witness, $\hat{W}$, 
has the following expansion: 
$\hat W = \sum_{\mu,\nu}w_{\mu\nu}\hat \sigma_{\mu}^A\otimes \hat \sigma_{\nu}^B$, where $\mu,\nu  = 0,1,2,3$ and $w_{\mu\nu}$ are real coefficients. We have  $\langle \hat{W}\rangle_{sep}\geq 0$, where $\langle \cdot \rangle_{sep}$ denotes an average for a separable state. 
We will show that with each  witness $\hat W$
one can 
associate  entanglement indicators 
for polarization measurements involving correlations of  Stokes observables for quantum optical fields.
The  maps are
$
\hat \sigma_{\mu}^A \otimes \hat{\sigma}_{\nu}^B  \to \hat S_{\mu}^A\hat S_{\nu}^B
$
and 
$
\hat \sigma_{\mu}^A \otimes  \hat{\sigma}_{\nu}^B  \to \hat \Theta_{\mu}^A\hat \Theta_{\nu}^B
$,
and they link $\hat{W}$ with its quantum optical analogues 
$
\hat{\mathcal{W}}_{S} = \sum_{\mu,\nu}w_{\mu\nu}\hat S_{\mu}^A\hat S_{\nu}^B,
$
and 
$
\hat{\mathcal{W}}_ {\Theta} = \sum_{\mu,\nu}w_{\mu\nu}\hat \Theta_{\mu}^A\hat \Theta_{\nu}^B,
$
which fulfill 
$\langle \hat{\mathcal{W}}_{S} \rangle_{sep}\geq 0$ 
and $\langle \hat{\mathcal{W}}_{\Theta} \rangle_{sep}\geq 0$. The proof goes as follows.

 %%%%%%%%%%%%%%%%%%%%%%%%%%%%%%%%%%%%%%%%%%%

{\it Normalized Stokes operators case.}---It is enough to prove that for any mixed state 
$\varrho$ 
%of two optical beams undergoing correlation 
%measurements of normalized Stokes parameters 
one can find a $4 \times 4$ 
 density matrix $\mathbf{\mathfrak{\hat R}}^{AB}_\varrho$ 
for a pair of qubits,  such that: 
\begin{equation}
\label{EQUALIZO1}
\frac{\langle{\hat{\mathcal{W}}_S}\rangle _\varrho}{\langle{\hat \Pi^A\hat \Pi^B}\rangle_\varrho}=\Tr \hat W \mathbf{\mathfrak{\hat R}}^{AB}_\varrho.
\end{equation}
First, we show  that  (\ref{EQUALIZO1}) holds for  any  pure state  $\ket{\psi^{AB}}$.
%, that is
%\begin{equation}
%\label{EQUALIZO}
 %\frac{\expval{\hat S_{\mu}^A\hat S_{\nu}^B}{\psi^{AB}}} {\expval{\hat \Pi^A\hat \Pi^B}{\psi^{AB}}}= \Tr \hat \sigma_{\mu}^A \hat \sigma_{\nu}^B \mathbf{\mathfrak{\hat R}}^{AB}_\psi.
%\end{equation} 

Let us denote the polarization basis $H$ and $V$ as
$\hat x_H = \hat x_1$ and $\hat x_V = \hat x_2$.
Normalized Stokes operators in arbitrary direction can be put  as
$\vec{m}\cdot \vec{S}^{X}$, where   
$\vec{m}$ is an arbitrary unit real vector,  or in the matrix form
$
\sum_{kl}\hat \Pi^X \frac{{ {\hat {x}}_k^{\dagger}}
(\vec{m}\cdot \vec{\sigma})_{kl}{{ \hat {x}_l}}}{\hat N^X}\hat \Pi^X,
$ with $\hat{x}=\hat{a}$ or $\hat{b}$ depending on the beam $X$, whereas 
  $\hat{S}_0^X$ reads 
$
\sum_{kl}\hat \Pi^X \frac{{ {\hat {x}}_k^{\dagger}} \delta_{kl}
{{ \hat {x}_l}}}{\hat N^X}\hat \Pi^X.
$
%The expectation value $\expval{\hat S_{\mu}^A\hat S_{\nu}^B}{\psi^{AB}}$ 
%can be re-expressed by the following steps. 
We introduce a set of states
\begin{eqnarray}
\label{super_psi}
\ket{\mathbf{\Psi}_{km}^{AB}} =
\hat{a}_k\hat{b}_m\frac{1}{\sqrt{\hat N^A\hat N^B}} \hat \Pi^A\hat \Pi^B \ket{\psi^{AB}},
\end{eqnarray}
where $k, m \in \{1,2 \}$. This allows us to put 
\begin{eqnarray}
\label{homomulti}
\expval{\hat S_{\mu}^{A} \hat S_{\nu}^B}{\psi^{AB}} 
 &=& \sum_{k,l=1}^2\sum_{m,n=1}^2\sigma^{kl}_{\mu }
\sigma^{mn}_{\nu }\bra{\mathbf{\Psi}^{AB}_{km}}\ket{
\mathbf{\Psi}^{AB}_{ln}} \nonumber \\
&=& \Tr\hat 
\sigma_{{\mu}}^A \otimes 
\hat \sigma_{{\nu}}^B
\hat R^{AB}_\psi, 
\end{eqnarray}
where the matrix elements of 
$\hat R^{AB}_\psi $ are $\bra{\mathbf{\Psi}^{AB}_{km}}\ket{\mathbf{\Psi}^{AB}_{ln}}$. 
As a Gramian matrix, ${\hat {R}}^{AB}_\psi$ is positive.  Except for  $|\psi^{AB}\rangle$ describing 
vacuum at one or both sides, we have 
$0<\Tr\hat R^{AB}_\psi= \langle \hat \Pi^A \hat \Pi^B \rangle \leq 1$. 
Thus, $\mathbf{\mathfrak{\hat R}}^{AB}_\psi=
\hat R^{AB}_\psi/\langle \hat \Pi^A \hat \Pi^B\rangle$ is an admissible 
density matrix of two qubits.

For mixed states $\varrho$, i.e., convex combinations of $\ket{\psi^{AB}_{\lambda}}$'s with weights $p_\lambda$, one gets 
$\hat{R}^{AB}_\varrho= \sum_{\lambda}p_{\lambda}{\hat{R}}^{AB}_{\lambda}$ which is positive definite, and its trace is
$ \sum_{\lambda}p_{\lambda}\Tr  \hat{R}^{AB}_{\lambda}\leq 1$. Thus after the re-normalization one gets a proper 
two-qubit density matrix $ \mathbf{\mathfrak{\hat R}}^{AB}_\varrho$.
As purity of a field state $\ket{\psi^{AB}_{\lambda}}$ does not warrant that the corresponding $\hat{R}^{AB}_{\lambda}$ is a projector, $ \mathbf{\mathfrak{\hat R}}^{AB}_\varrho$ does not have to have the same convex expansion coefficients  in terms of pure two-qubit states, as $\varrho$ in terms of $\ket{\psi^{AB}_{\lambda}}$'s.

For any separable pure state of two optical beams 
$|\psi^{AB}\rangle_{prod}$, defined 
as $F^\dagger_A F^\dagger_B|\Omega\rangle$, 
where $F^\dagger_X$ is a polynomial function of  
creation operators 
 %creating a superposition of Fock  states 
for beam (modes) $X$, and $|\Omega\rangle$ is the vacuum state of both beams,
the matrix $\hat R^{AB}$ factorizes: 
$\hat R^{AB} = \hat R^{A}
\hat R^{B}$.  
Simply,  ${}_{prod}\bra{\mathbf{\Psi}^{AB}_{km}}\ket{\mathbf{\Psi}^{AB}_{ln}}_{prod}$ factorizes to 
$
\bra{\mathbf{\Psi}^{A}_{k}}\ket{\mathbf{\Psi}^{A}_{l}} \bra{\mathbf{\Psi}^{B}_{m}}\ket{\mathbf{\Psi}^{B}_{n}}, 
$
where 
$\bra{\mathbf{\Psi}^{X}_{k}}\ket{\mathbf{\Psi}^{X}_{l}}$ 
are elements  of matrix $\hat R^{X}$ 
and $
|\mathbf{\Psi}_{l}^{X}\rangle =  
\hat{x}_l\frac{1}{\sqrt{\hat N^X}} \hat{\Pi}^X F^{\dagger}_X|\Omega\rangle.
$
As 
$\langle \Omega|F_X \hat{\Pi}^XF^{\dagger}_X|\Omega\rangle^{-1}\hat R^{X}$ 
can be shown to be 
a qubit density matrix and $\langle\hat{W}\rangle_{sep}\geq 0$, 
therefore for pure separable  states of the optical 
beams $\langle\hat{\mathcal{W}}_S\rangle_{prod}\geq 0 $. Obviously, 
$\langle \mathcal{\hat W}_S \rangle_{sep}\geq 0 $ also for all mixed separable states.

%%%%%%%%%%%%%%%%%%%%%%%%%%%%%%%%

{\it Standard Stokes operators case.}--- Any standard Stokes operator can be put as
$\vec{m} \cdot \vec{\Theta}^X = 
 \sum_{kl}
 \hat{x}_k^{\dagger}
(\vec{m}\cdot \vec{\sigma})_{kl}
 \hat{x}_l.
$
We introduce  state vectors 
$\ket{\mathbf{\Phi}_{jk}^{AB}} = \hat a_{j}\hat b_k  \ket{\psi^{AB}}$.
One has 
\begin{eqnarray}
\expval{\hat \Theta_{\mu}^A \hat \Theta_{\nu}^B}{\psi^{AB}} = \Tr \hat \sigma^A_{\mu}
\hat \sigma^B_{\nu} \hat P^{AB},   
\end{eqnarray}
where the matrix $\hat P^{AB}$ 
has entries $\bra{\mathbf{\Phi}_{km}^{AB}}\ket{\mathbf{\Phi}_{ln}^{AB}}$, 
it
is {positive definite}, 
and its trace is $ \langle \hat N_A \hat N_B \rangle.$
Thus, $\mathbf{\mathfrak{\hat P}}^{AB}  = \hat P^{AB}/\langle \hat N^A \hat N^B\rangle$ is an admissible two-qubit density matrix, and one has
$
\langle \hat{\mathcal{W}}_\Theta\rangle_\varrho/\langle \hat N ^A\hat N^B\rangle_\varrho = \Tr \hat W
\mathbf{\mathfrak{\hat P}}^{AB}_\varrho.
$
All that leads to $\langle \hat{\mathcal{W}}_\Theta\rangle_{sep}\geq 0$.
Note that, for a general 
state $\mathbf{\mathfrak{\hat R}}^{AB}_\varrho$ 
does not have to be equal to $ \mathbf{\mathfrak{\hat P}}^{AB}_\varrho$. Still,
$\mathbf{\mathfrak{\hat R}}^{AB} = \mathbf{\mathfrak{\hat P}}^{AB}$ for states of defined photon numbers in both beams.

{\it Reverse map.}---
Any linear  separability condition  expressible in  terms of correlation functions of normalized  Stokes Parameters 
reads:
$
\sum_{\mu\nu} \omega_{\mu\nu}\langle \hat{S}^A_\mu \hat{S}^B_\nu\rangle_{sep} \geq 0.
$
%This must hold for any separable  state of the optical field.
As two-photon states, with one at A and the other at B, are  possible field states, thus for any separable such state we must have 
$
\sum_{\mu\nu} \omega_{\mu\nu}\langle \hat{S}^A_\mu \hat{S}^B_\nu\rangle_{sep-2-ph} \geq 0.
$
This is algebraically equivalent to
$
\sum_{\mu\nu} \omega_{\mu\nu}\langle \hat{\sigma}_\mu \otimes \hat{\sigma}_\nu\rangle_{sep} \geq 0,
$
for any two-qubit state. We get an entanglement witness. Therefore, we have an isomorphism. Similar proof applies to standard Stokes observables.

{\it Examples.}---In the Supplemental Material~\cite{supp}, 
we show some examples of entanglement indicators which can be derived with the above method. 
This includes a necessary and sufficient conditions for detection of entanglement of two optical beams with correlations of Stokes parameters of the two considered  kinds. 
%We also show that detection of entanglement  with  normalized Stokes parameters can be more resistant with respect to noise than with standard ones.

%Let us present an important example.
%%%%%%%%%%%%%%%%%%%%%%%%%%%%%%%%%%%%%%%%%%%

{\it Detection losses.}---
%The 
%entanglement conditions  for optical fields, 
%defined by $\hat{\mathcal{W}}_S$ and $\hat{\mathcal{W}}_\Theta$,  
%are highly resistant to collection-detection losses. 
Consider the usual model of losses:
a perfect detector in front of which is a beamsplitter of transmission amplitude ${\eta}$, with the reflection channel describing the losses. 
Then, 
$\langle \hat \Theta_{\mu}^A\hat \Theta_{\nu}^B\rangle$ 
scales down as $\eta^A\eta^B$ (see Sec. II in the Supplemental Material~\cite{supp}), where 
$\eta^X$ for $X=A,B$ is the local detection efficiency.
We have a full resistance of entanglement 
detection,  using any $\hat{\mathcal{W}}_\Theta$, with respect to such losses.
A different character of 
losses may lead to threshold efficiencies.

For  the normalized Stokes parameters,  it is enough to consider only pure states, 
because mixed ones, as convex combinations of such, 
cannot introduce anything new in 
entanglement  conditions linear with respect to the density matrix. 
Any pure state is a superposition of Fock states $|F\rangle= |n_{i}^{A},n_{i_\perp}^{A}, n_{j}^{B},n_{j_\perp}^{B} \rangle$, where $n_{i}^{X}$ denotes the number of $i$ polarized photons in beam $X$, and  $\hat{S}_{\mu}^A\hat S_{\nu}^B$ are diagonal with respect to the Fock basis related with them. Thus, the dependence on efficiencies  of the  value of an entanglement   indicator, in the case of a pure state, depends on the behavior of its Fock components. One can show, see Sec. II in the Supplemental Material~\cite{supp},
%~\cite{supp_loss} 
that 
$
%\begin{equation}
\label{EFFICIENCY}
 {\langle F_\eta|\hat {S}_{\mu}^A\hat{ S}_{\nu}^B |F_\eta  \rangle} = H_F
{\langle F|\hat {S}_{\mu}^A\hat{ S}_{\nu}^B |F\rangle},
%\end{equation}
$
where $|F_\eta\rangle$ is the state $|F\rangle$ after the above described losses in both channels, and 
$H_F=\langle F_\eta| \hat{S}_{0}^A\hat S_{0}^B |F_\eta \rangle $, which reads $\prod_{X=A,B}[1-(1-\eta^X)^{m^X}]$, where  $m^X$ 
is the total number of photons in channel $X$, before the losses. 
Expanding  $|F\rangle$ in terms of Fock states with respect to different polarizations than $i,i_\perp$ 
and $j,j_\perp$, does not change the values of $m^X$, and 
thus the formula stays put for any indices.
%$\nu$ and $\mu$.
%\textcolor{blue}{Note that the  values of 
%$m^X$ remain unchanged regarding    
%expanding of $|F\rangle$ in terms of Fock states 
%with respect to any different polarizations   
%} 
Again we have a strong resistance of the 
entanglement indicators with respect to losses. Especially for states with high photon numbers, the entanglement conditions based on normalized Stokes
 parameters, may be more resistant to losses, because  $0<\eta < 1$ one has $\eta<1-(1-\eta)^n$.

% %Thus $H_F$, with growing $n$, quickly approaches $1$. %
%Generally, the magnitude of violation of the zero  
%threshold would depend on $\eta$'s
%in a complicated way, 
%dependent on photon number distribution in the state before losses.

%%%%%%%%%%%%%%%%%%%%%%%%%%%%%%%%%%%%%%%%%%%%%%%%%%%%%%%%
{\it Multi-party case.}---  Consider three parties, and the case of indicators of genuine three-beam entanglement. Any genuine three-qubit entanglement witness $\hat{W}^{(3)}$ has the property that it is positive for  pure  product three-qubit states   $|\xi\rangle_{AB,C}=|\psi\rangle_{AB}|\phi\rangle_C$,   for similar ones with qubits   permuted, and  for all convex combinations of such states.  With any pure partial product state of the optical beams, e.g. $|\Xi\rangle_{AB,C}=F^\dagger _{AB} F^\dagger_C|\Omega\rangle $, where $F^\dagger_{AB}$ is an operator built of creation operators for beams $A$ and $B$, etc.,  one can associate, in a similar way as above, 
a partially factorizable three-qubit density matrix $\mathfrak{\hat{R}}^{AB}_\psi\mathfrak{\hat{R}}^{C}_\phi$.
Thus, the homomorphism  works. Generalizations  are obvious. 
%All witnesses of ``genuine'' multi-qubit entanglement can be translated into indicators of genuine multi-beam entanglement. 

%%%%%%%%%%%%%%%%%%%%%%%%%%%%%%%%%%%%%%%%%%%%%%%%

{\it  General Theory.}---Consider a  beam of $d_A$ quantum 
optical modes propagating toward a
measuring station $A$, and a beam of $d_B$ modes toward station $B$. We associate with 
the situation  a $d_A \times d_B$ dimensional Hilbert Space, $\mathbb{C}^{d_A} \otimes \mathbb{C}^{d_B}$, which contains pure states of a pair of qudits
of dimensions $d_A$ and $d_B$.  
For $X=A,B$, let $\hat V_i^X$, with $i=1,...,d_X^2$, be an orthonormal, i.e.  $\Tr \hat V_i^X \hat V_j^X=\delta_{ij}$, Hermitian  basis of the space of Hermitian operators acting on $\mathbb{C}^{d_X}$. 
Therefore, products $\hat V_i^A\otimes \hat V_j^B$ form an orthonormal basis of the space of Hermitian operators acting on $\mathbb{C}^{ d_A}\otimes \mathbb{C}^{d_B}$.
 Thus, any entanglement witness for the pair of qudits,  $\hat W$, can be expanded into 
\begin{equation}\label{EXPANSION}
\hat W=\sum_{j=1}^{d_A^2}\sum_{k=1}^{d_B^2}w_{jk}\hat V_j^A\otimes 
\hat V_k^B,
\end{equation}
with real  $w_{jk}$.  The optimal
expansion (with the minimal number of terms) is to use a Schmidt basis for $\hat{W}$. % (if we assume $d_A\leq d_B $, we would have only $d_A^2$ terms, and thus much less pairs of local measurements to get  $\langle \hat{W}\rangle $).

Each $\hat V^X_j$
can be decomposed to a linear combination of
its spectral projections linked with their respective eigenbases, $|x^{(j)}_l\rangle$, 
where $x=a$ or $b$ 
consistently with $X$  and $l=1,..., d_X$. If one fixes a certain pair of bases in
 $\mathbb{C}^{d_A}$ and $\mathbb{C}^{d_B}$ as ``computational ones", i.e., starting ones, denoted  as $|l_x\rangle$, one can always find  local unitary matrices  $U^X(j)$ such that $U^X(j)|l_x\rangle= |x^{(j)}_l\rangle$. The construction of Reck et al.~\cite{RECK} fixes (phases in) a local multiport interferometer, which performs such a transformation. We shall call such interferometers $U^X(j)$ ones.
In the case of field modes  a passive interferometer performs the following mode transformation: 
$\sum_{k} U^X(j)_{lk}\hat x^\dagger_k= \hat x^\dagger_{l}(j)$, where $\hat x^\dagger_{l}(j)$ is the photon creation operator in the $l$-th exit mode of  interferometer $U^X(j)$. 
 
A two-party entanglement witness $\hat {\mathcal{W}}_R$ 
for optical fields, 
which uses  correlations of intensity {\it rates} 
behind  pairs  of  $U^X(j)$ interferometers can be constructed as follows.
For the output $l_x$  of an interferometer, one defines rate observables
 as
 $\hat{r}_{l_x}=\hat{\Pi}^X\frac{\hat n_{l_x}}{\hat{N}^X} \hat{\Pi}^X$,  where $\hat{N}^X=\sum_{l_x=1}^{d_X}\hat{n}_{l_x}$. % is sum of photon number operators for all outputs of the interferometer.
The witness $\hat W$ 
expanded  in terms of the computational basis:
\begin{equation} \label{KETBRA}
\hat W=\sum_{k,m}^{{d}_{A}}\sum_{l,n}^{{d}_{B}} w_{klmn}|k_a,l_b\rangle\langle m_a,n_b|,
\end{equation}
allows us to form an entanglement witness for  fields: 
\begin{equation}
\label{RATE_WITNESS}
\hat {\mathcal{W}}_R=\sum_{k,m}\sum_{l,n}w_{klmn}
\hat \Pi^A\hat \Pi^B\frac { \hat a^\dagger_k \hat b^\dagger_l \hat a_m \hat b_n} 
{\hat N^A \hat N^B}\hat \Pi^A\hat \Pi^B.
\end{equation}
For any pure state of the quantum beams $|\Psi\rangle $ 
\begin{equation}
\frac{\langle \Psi |\hat{\mathcal{W}}_R|\Psi\rangle}{ \langle \Psi|\hat \Pi^A\hat \Pi^B|\Psi\rangle} = 
\Tr \hat W \hat{ \mathcal{R}},
\label{eq:avg_rate}
\end{equation}
where the matrix $\hat{\mathcal R}$ has elements $r_{klmn}$
\begin{equation}
 r_{klmn} =  \frac{1}{\langle \Psi|\hat \Pi^A\hat \Pi^B|\Psi\rangle}
 \langle \Psi |\hat \Pi^A\hat \Pi^B\frac { \hat a^\dagger_k \hat b^\dagger_l \hat a_m \hat b_n} 
 {\hat N^A \hat N^B}\hat \Pi^A\hat \Pi^B|\Psi\rangle.
\end{equation} 
Using a   
generalization of the earlier derivations  
one can show that   $\hat{\mathcal R}$  
is a  two-qudit density matrix, and so on.

The actual measurements, 
to be correlations of local ones, should 
be performed using the 
sequence of pairs of $U^X(j)$ interferometers, 
which enter the expansion of the two-qudit entanglement witness (\ref{EXPANSION}).  In the entanglement 
indicator the rates  at   
output $x_l(j)$ of the given local 
interferometer $U^X(j)$ are multiplied by the respective  
eigenvalue of $\hat V_j^X$ related 
with the eigenstate $|x^{(j)}_l\rangle$. 

To get an entanglement witness for intensities $\mathcal{\hat{W}}_I$  we take  $\hat W$  and replace the computational basis kets and bras  by suitable creation and annihilation operators:
\begin{equation}
\label{INTENSE_WITNESS}
\hat {\mathcal{W}}_{I}=\sum_{k,m}^{{d}_{A}}\sum_{l,n}^{{d}_B}w_{klmn} 
\hat a^\dagger_k \hat b^\dagger_l \hat a_m \hat b_n.
\end{equation}
For any pure state of the quantum beams $|\Psi\rangle$ one has
$
\frac{\langle \Psi|\hat {\mathcal{W}}_{I}|\Psi\rangle}{ \langle \Psi|\hat N_A \hat N_B|\Psi\rangle} = \Tr \hat W {\hat{\mathcal P}},
$
where the matrix $\hat{\mathcal P}$ has elements
$
\frac{1}{\langle \Psi| \hat N_A\hat N_B|\Psi\rangle}\langle \Psi | \hat a^\dagger_k \hat b^\dagger_l \hat a_m \hat b_n | \Psi\rangle,
$
and has all properties of a  two-qudit density matrix.

{\it Example showing further extension to unitary operator bases.}---Let $d$ be a power of a prime number.
Consider $d_A=d_B=d$ beams experiment (see Fig.~\ref{fig:exp}),  with  families of $U^X(m)$ 
interferometers which link the computational basis of a qudit  
with an unbiased basis $m$, belonging to the full set of $d+1$ mutually unbiased ones~\cite{WOOTERS, MUBs}.
%To get a homomorphic 
%relation between a two-qudit density matrix 
%and matrices $\hat{\mathcal R}$  and $\hat{\mathcal P}$ 
%for the fields 
We introduce a set of unitary observables 
for a qudit:
$
%\begin{eqnarray}
\hat{q}_k(m)= \sum_{j=1}^d\omega^{jk}|j(m)\rangle \langle j(m)|,
\label{EQ:UNI_OBS}
%\end{eqnarray}
$
with $ | j(m)\rangle =U(m)|j\rangle$ and it is the $j$-th member of $m$-th mutually unbiased basis, and $\omega=\exp(2\pi i/d)$. 
Operators $\hat{q}_k(m)/\sqrt{d}$ with $k = 1,...,d-1$ 
and $m=0,...,d$ and $\hat{q}_0(0)/\sqrt{d}$ form an orthonormal basis in the 
Hilbert-Schmidt space of all $d\times d$ matrices (see Sec. III in the Supplemental Material~\cite{supp}).
Thus, we can expand any  qudit density matrix as
\begin{equation}\label{COMPLETE}
{\varrho} =\frac{1}{\sqrt{d}} \left[ 
c_{0,0}\hat{q}_0(0) + 
\sum_{m=0}^d\sum_{k=1}^{d-1}c_{m,k}\hat{q}_k(m) \right],
\end{equation}
where 
$
c_{m,k} = \Tr \hat{q}_k^{\dagger}(m) \varrho / \sqrt{d},
$
and $c_{0,0}=1/\sqrt{d}$.
As the basis observables 
are unitary the expansion coefficients of an entanglement witness operator in terms of such tensor products of such  bases
are in general complex. 
This is no problem for theory, 
but renders useless a direct application in experiments, 
as one cannot expect the experimental averages to be real, and thus one has to introduce modifications. Below we present one.

The condition  $\Tr \varrho^2\leq 1$  
can be put as
\begin{equation}
\frac{1}{d}+\frac{1}{d}\sum_{m=0}^d\sum_{k=1}^{d-1}
\left| \Tr  \varrho \hat{q}_{k}(m) \right|^2  \leq 1.
\label{SINGLE2} 
\end{equation}
Thus, applying Cauchy-Schwartz estimate,
we get immediately a separability condition for two qudits: 
\begin{equation}
\sum_{m=0}^d\sum_{k=1}^{d-1}
 \left| \Tr  \varrho^{AB}_{sep} 
\hat{q}^{A}_{k}(m)\hat{q}^{B\dagger}_{k}(m) \right|  \leq (d-1).
\label{PAIR2}
\end{equation}
Our general method defines 
a Cauchy-Schwartz-like separability condition homomorphic with (\ref{PAIR2}) as
\begin{equation}
\label{MULTIWARUNEKnew}
\sum_{m=0}^d \sum_{k=1}^{d-1}|\langle \hat Q_k^A(m)\hat 
Q_k^{B{\dagger}}(m)\rangle_{sep}| \leq
(d-1)\langle 
 \hat \Pi^A \hat\Pi^B\rangle_{sep},
\end{equation}
where
\begin{equation}
\label{MULTIOBS}
\hat Q_{k}^X(m) 
= \sum_{j=1}^d \hat \Pi^{X}\frac{\omega^{jk}\hat n^X_j(m)}{\hat N^{X}}\hat \Pi^{X}.
\end{equation}
Here $\hat n^X_j(m) = \hat x^{\dagger}_j(m)\hat x_j(m)$ is a photon number operator for  output mode $j$ of a multiport $m$, at station $X$. 
For generalized observables based on intensity, one can introduce
$
\label{MULTIOLD}
{\hat \chi}_{k}(m) = \sum_{j=1}^d\omega^{jk} \hat n_j(m)
$
to get the following separability condition:
\begin{equation}
\label{MULTIWARUNEKold}
\sum_{m=0}^d \sum_{k=1}^{d-1} 
|\langle{ \hat \chi }_k^A(m){\hat 
\chi}_k^{B{\dagger}}(m)\rangle_{sep}| \leq
(d-1)\langle 
 \hat N^A \hat N^B\rangle_{sep}.  
\end{equation}
Supplemental Material presents other examples~\cite{supp}.

%In Appendix we give an example showing that the indicator based on  observables employing rates  (\ref{MULTIWARUNEKnew}) for $d^2$ mode  squeezed vacuum allows more noise resistant entanglement detection  than
%(\ref{MULTIWARUNEKold}).
%Note also,  that in case of $d=2$ the multport  observables  reduce to  normalized and  standard  Stokes 
%parameters, respectively. 

%The interaction in the source $S$, given by the Hamiltonian (6), generates a d-mode bright squeezed vacuum state (8). This entangled state leads to perfect EPR correlations between the local conjugate modes.

{\it Implications for optical coherence theory.}---The approach can be generalized further.
Let us take as an example Glauber's correlation functions for optical fields, say $G^{(4)}$
in the form of 
$
\langle \hat{I}_A(\vec{x},t) \hat{I}_B(\vec{x}',t')\rangle
$,
where the intensity operator has the usual form of $I_X(\vec{x}, t)=\hat{F}_X^\dagger(\vec x, t) \hat{F}_X(\vec x, t)$,
with normal ordering requiring that  operator $\hat{F}_X(\vec x, t)$ is built out of local annihilation operators. The idea of normalized Stokes operators
suggests the following alternative correlation function $\Gamma^{4}(\vec{x},t; \vec{x}', t')$ given by
\begin{eqnarray}
\label{GLAUBER}
\langle \Pi^A\Pi^B \frac{\hat{I}_A(\vec{x},t) \hat{I}_B(\vec{x}',t')}{\int_{a(A)} d\sigma(\vec{x}) \hat{I}_A(\vec{x},t)\int_{a(B)} d\sigma(\vec{x}') \hat{I}_B(\vec{x}',t')}\Pi^A\Pi^B\rangle,&& \nonumber \\
\end{eqnarray}
where $a(X)$ denotes the overall aperture of the detectors in location $X$. Obviously one has $\int_{a(A)} d\sigma(\vec{x})\int_{a(B)} d\sigma(\vec{x}') \Gamma^4(\vec{x},t; \vec{x}', t')=\langle\Pi^A\Pi^B\rangle$, and for fixed $t$ and $t'$ one can define 
\begin{eqnarray*}
&&\varrho(\vec{x},\vec{y},\vec{x}' \vec{y}')_{t,t'}
= {\langle\Pi^A\Pi^B\rangle}^{-1} \nonumber \\
&&\times \langle \Pi^A\Pi^B
 \frac{\hat{F}^\dagger_A(\vec{y},t)\hat{F}_A(\vec{x},t)
\hat{F}^\dagger_B(\vec{y}',t') \hat{F}_B(\vec{x}',t')}
{\int_{a(A)} d\sigma(\vec{x}) \hat{I}_A(\vec{x},t)\int_{a(B)} d\sigma(\vec{x}') \hat{I}_B(\vec{x}',t')}\Pi^A\Pi^B\rangle,
\end{eqnarray*}
which behaves like a proper two-particle density matrix, provided one constrains the range of $\vec{x}, \vec{y}, \vec{x}', \vec{y}'$ to appropriate sets of apertures. As our earlier considerations use simplified forms of (\ref{GLAUBER}), it is evident that such correlation functions may help us to unveil non-classicality in situations
in which the standard ones fail, see e.g.~\cite{ROSOLEK}.

{\it Bell inequalities.}---The above ideas allow one to  introduce a general mapping of qudit Bell inequalities
to the ones for optical fields.
A two-qudit Bell inequality for a final number of local measurement settings $\alpha$ and $\beta$ has the following form:
\begin{eqnarray}
\label{BELL}
&&\sum_{\alpha\beta}\sum_{i=1}^{d_A}\sum_{j=1}^{d_B} K^{ij}_{\alpha\beta} P_{ij}(\alpha, \beta) \nonumber \\
&&+ \sum_{i=1}^{d_A}\sum_\alpha N^i_\alpha P_i(\alpha) 
+ \sum_{j=1}^{d_B}\sum _\beta M^j_\beta P_j(\beta)\leq L_R,
\end{eqnarray}
where $P_{ij}(\alpha, \beta)$ denotes the probability of the qudits ending up respectively at detectors $i$ and $j$, when the local setting are as indicated, and $\sum_jP_{ij}(\alpha, \beta)=P_{i}(\alpha)$ and  $P_{j}(\beta)=\sum_iP_{ij}(\alpha, \beta)$.
 The coefficient matrices $K, N, M$ are real, and $L_R$ is the maximum value allowed by local realism. The bound is calculated by putting $P_{ij}(\alpha,\beta)= D^i(\alpha)D^j(\beta)$ and $P_{i}(\alpha)= D^i(\alpha)$,  
$P_{j}(\beta)= D^j(\beta)$, with constraints  $0\leq D^i(\alpha/\beta )\leq1$, and    $\sum_{i=1}^{d_{A/B}}D^i(\alpha/\beta)=1$. As for a given run of a quantum optical experiment local measured photon intensity rates $r_i(\alpha)$ and $r_j(\beta)$ satisfy exactly the same constraints.
%the above general Bell inequality holds for any local realistic values of these. 
We can  replace $P_{ij}(\alpha, \beta)\rightarrow \langle r_i(\alpha)r_j(\beta)\rangle_{LR}$, and $P_{i}(\alpha)\rightarrow \langle r_i(\alpha)\rangle_{LR}$, etc., where $\langle.\rangle_{LR}$ is an average in the case of local realism. The bound $L_R$ stays put. To get a Bell operator we further replace the above by rate observables
$\hat{r}_i(\alpha) \hat{r}_j(\beta)$, etc. Thus any (multiparty)  Bell inequality,  see e.g.~\cite{BELL-RMP}, can be useful in quantum optical intensity  (rates) correlation experiments.
The presented methods for entanglement indicators and Bell inequalities allow also to get steering inequalities for quantum optics.

{\it Conclusions.}---We present tools for a construction of entanglement indicators for optical fields, inspired by the vast literature  \cite{HORODECKIS} on entanglement witnesses for finite dimensional quantum systems.
The indicators  would be handy for more intense light beams in states of undefined photon numbers, especially  in the  emerging field of  integrated optics multi-spatial mode interferometry (see Supplemental Material~\cite{supp} for examples). One may expect applications in the case of many-body systems, e.g. for an  analysis of non-classicality of correlations in Bose-Einstein condensates, like in the ones reported in \cite{SCHMIED}.

\begin{acknowledgments}
{\it Acknowledgments.}---The work is part of the ICTQT IRAP project of FNP, financed by structural funds of EU. MZ acknowledges COPERNICUS grant-award, and discussions with profs. Maria Chekhova and Harald Weinfurter. JR acknowledges the National Research Foundation, Prime Minister’s Office, Singapore and the Ministry of Education, Singapore under the Research Centres of Excellence programme, and discussions with prof. Dagomir Kaszlikowski. MW acknowledges NCN grants number 2015/19/B/ST2/01999 and 2017/26/E/ST2/01008.
\end{acknowledgments}

\section*{Supplemental material}

We give here several examples, and more details concerning some derivations.   All separability conditions are generalizations or tighter versions of conditions
presented in \cite{BOUW-2, BOUW, zuko, MULTIPORT, MULTIPORT2}, which were derived using various less general approaches.

\section{Necessary and Sufficient conditions for intensity and rate correlations to reveal entanglement}
\label{apx:exp_iff_qubit}
Two-qubit states are separable 
if and only if their partial transposes are positive. 
Yu {\it et al.} derived an equivalent family of conditions for two-qubit states~\cite{CHINY} in a form of an inequality, which reads 
\begin{eqnarray}
\label{CHINSKIPAULI}
\langle \hat \sigma_{1}^{A}\hat \sigma_{1}^{B} +  \hat \sigma_{2}^{A}\hat \sigma_{2}^{B} \rangle^2  &+& \langle \hat \sigma_{3}^{A}\hat{\sigma}_0^{{B}} + \hat{\sigma}_0^{{A}} \hat \sigma_{3}^{B}\rangle^2 \nonumber \\
 &\leq& \langle \hat{\sigma}_0^{A} 
 \hat{\sigma}_0^{B} + \hat \sigma_{3}^{A}\hat \sigma_{3}^{B} \rangle^2,
 \end{eqnarray}
where $\hat{\sigma}^X_j=\vec{n}^X_j\cdot \vec{\sigma}^X$ for $X=A, B$, and the unit vectors $\vec{n}^X_j$ form a right-handed  Cartesian basis triad. 
If a two-qubit state is entangled, 
then there exists at least one pair 
of such triads 
for which the inequality is violated.
The conditions can be put in a form of a 
family of  entanglement witnesses:
\begin{eqnarray}
\label{ENT-WIT-YU}
 W(\alpha) =\hat{\sigma}_0^{A}\hat{\sigma}_0^{B} + \hat \sigma_{3}^{A}\hat \sigma_{3}^{B} &+& \sin{\alpha}( \hat \sigma_{1}^{A}\hat \sigma_{1}^{B} + \hat \sigma_{2}^{A}\hat \sigma_{2}^{B}) \nonumber \\   
&+&  \cos{\alpha}( \hat \sigma_{3}^{A}\hat{\sigma}_0^{{B}} + \hat{\sigma}_0^{{A}} \hat \sigma_{3}^{B}).
 \end{eqnarray}
Our homomorphisms can be used to get the following~\cite{zuko1}: for normalized Stokes operators
\begin{eqnarray}
\label{CHINSKISTOKES1}
\langle \hat S_{1}^{A}\hat S_{1}^{B} + \hat S_{2}^{A}\hat S_{2}^{B} \rangle^2 &+& \langle \hat S_{3}^{A}\hat \Pi^{{B}} + 
\hat \Pi^{{A}}\hat S_{3}^{B}\rangle^2 \nonumber \\
&\leq&  \langle \hat \Pi^{A}\hat \Pi^{B} + \hat S_{3}^{A}\hat S_{3}^{B}\rangle^2,
\end{eqnarray}
and for standard ones
\begin{eqnarray}
\label{CHINSKISTOKES2}
\langle \hat \Theta_{1}^{A}\hat \Theta_{1}^{B} + \hat \Theta_{2}^{A}\hat \Theta_{2}^{B} \rangle^2  &+& \langle \hat \Theta_{3}^{A}\hat N^{{B}} + \hat N^{{A}}\hat \Theta_{3}^{B}\rangle^2 \nonumber \\
&\leq& \langle \hat N^{A}\hat N^{B} + \hat \Theta_{3}^{A}\hat \Theta_{3}^{B}\rangle^2.
\end{eqnarray}
The homomorphisms warrant that the violations of conditions~(\ref{CHINSKISTOKES1}) and~(\ref{CHINSKISTOKES2}) are {\it necessary and sufficient} to detect entanglement via measurements of correlations of the Stokes observables.
%of the kind which they involve.
That is, any other condition is sub-optimal, including the ones presented in \cite{BOUW}, \cite{Iskhakov12} and \cite{BOUW-2} for standard Stokes observables.

From the necessary and sufficient condition (\ref{CHINSKISTOKES1}) one can derive its corollary, which is a necessary condition for separability:
\begin{equation}
\label{WARUJNOWYauto} 
\sum_{j=1}^{3} |\langle{\hat S}_{{j}}^{A}
{ \hat S}_{{j}}^{B} \rangle_{sep}|  \leq  \langle \hat 
\Pi^A
\hat \Pi^B \rangle_{sep}.
\end{equation}
The condition can be thought as a more tight refinement of the result in \cite{BOUW-2}.
It can be derived using the fact that for two qubits any of  the observables
$\sum_k s_k \sigma^A_k \sigma^B_k+ \sigma^A_0 \sigma^B_0$, 
for arbitrary $s_k=\pm1$ is  non-negative for separable states.  This can be reached via an application of the Cauchy inequality for a product  pure states of a pair of qubits.  Next we apply the homomorphism.
One can also see that  (\ref{WARUJNOWYauto}) is  the separability condition (14) in the main text for $d=2$. 

For the standard Stokes operators the associated separability condition  (\ref{WARUJNOWYauto}) reads
\begin{equation}
\label{WARUJSTARYauto} 
\sum_{j = 1}^{3} |\langle\hat  \Theta_{j}^{A} \hat \Theta_{j}^{B} \rangle_{sep}|  \leq  \langle 
{\hat N}^{A} {\hat N}^{B} \rangle_{sep}. 
\end{equation} 
This is a tighter version of the condition given in \cite{BOUW-2}.

For states, which locally lead to vanishing averages of local Stokes parameters, here  $\langle \hat{S}^A_i \hat{\Pi}^B\rangle =0 $, etc., 
(e.g., for  an ideal  four-mode bright squeezed vacuum, see below), 
the conditions (\ref{WARUJNOWYauto}) and  (\ref{CHINSKISTOKES1}) are equivalent.
Thus, in such a case  the Cauchy inequality based 
condition is  necessary and sufficient for detection of entanglement with normalized Stokes operators. 
A similar statement can be produced for the analog condition involving
traditional Stokes parameters $\Theta_j$, given by (\ref{WARUJSTARYauto}).

{\it Cauchy-like inequality condition vs. EPR inspired approach}.---Consider four-mode (bright) squeezed vacuum represented by
\begin{equation}
\label{BOH_PROJ_0}
\ket{\Psi^-}   =  
\frac{1}{\cosh^2\Gamma} \sum_{n=0}^{\infty}\sqrt{n+1} \tanh^n\Gamma
\textcolor{black}{\ket{\psi^{n}_-}},  
\end{equation}
where $\Gamma$ describes a gain which is proportional to the pump power, and 
$\ket{\psi^{n}_{-}}$ reads
\begin{equation}
 \label{PSINONOISE}
 \ket{\psi^n_-} = \frac{1}{n! \sqrt{n+1}}
\bigg(\hat a_i^{\dagger}\hat 
b_{i_{\perp}}^{\dagger} -\hat 
a_{i_{\perp}}^{\dagger}
b_i^{\dagger} \bigg)^n 
\ket{\Omega},
\end{equation} 
where $\ket{\Omega}$ is the vacuum state.

Perfect EPR-type anti-correlations of which are the main trait of the  state allow one
to formulate the following appealing separability 
condition (Simon and Bouwmeester, \cite{BOUW}): 
\begin{equation} 
\label{OLDANDWRONG}
 \sum_{j=1}^3 \langle (\hat \Theta_{j}^A 
+\hat \Theta_{j}^B )^2\rangle_{sep} \geq 
2(\langle  \hat N^{A}\rangle  + \langle \hat 
N^{B}\rangle)_{sep}.
\end{equation}
Note, that for $|\Psi^-\rangle$ and each $|\psi^n_-\rangle$ the left-hand side (LHS) of the above is vanishing.

The underlying inequality beyond the condition (\ref{OLDANDWRONG}) can be extracted with the use of  
well-known operator identity (see e.g. \cite{KLYSHKO}): 
\begin{equation} 
\label{RELOLD}
\sum_{j=1}^3 \hat  \Theta_j^2 = \hat 
N(\hat N+2).
\end{equation}
Using this the (\ref{OLDANDWRONG}) boils down to
\begin{equation}
\label{WARUJ}
- \sum_{j=1}^3 \langle \hat \Theta_{j}^A
\hat \Theta_{j}^B\rangle_{sep} \leq \frac{1}{2}\langle \hat
N_{A}^2\rangle_{sep}
+ \frac{1}{2}\langle  \hat N_{B}^2\rangle_{sep},
\end{equation}
which by the way can be generalized to 
$
2 \sum_{j=1}^3| \langle \hat \Theta_{j}^A
\hat \Theta_{j}^B\rangle_{sep}| \leq\langle \hat
N_{A}^2\rangle_{sep}
+ \langle  \hat N_{B}^2\rangle_{sep}.
$

Simon-Bouwmeester EPR-like condition (\ref{OLDANDWRONG}), or equivalently  (\ref{WARUJ}), cannot be 
considered as an entanglement indicator for fields 
$\hat{\mathcal{W}}_{\Theta}$ 
homomorphic in the way proposed here, with a two-qubit (linear) entanglement witness $\hat{W}$. 
Detection of entanglement with (\ref{OLDANDWRONG}) depends on a detector efficiency. 
The threshold  efficiency for  entanglement detection, in the case of $2 \times 2$ mode squeezed vacuum $|\Psi^-\rangle$ in (\ref{BOH_PROJ_0}),
considered in \cite{BOUW}
is given by $\eta_{crit} = 1/3$.
It does not depend on the gain parameter  $\Gamma$. 
Obviously,  as $\langle \hat{N}_A \hat{N}_B\rangle  \leq  \frac{1}{2}\langle\hat
N_{A}^2\rangle 
+ \frac{1}{2}\langle  \hat N_{B}^2\rangle,$ the inequality (\ref{WARUJ}) is not 
optimal.
A more optimal option is to estimate from below the LHS of (\ref{WARUJ}) using a corollary of the 
Cauchy-like inequality 
$
\label{WARUJCS}
-\sum_{j=1}^3 \langle \hat \Theta_{j}^A
\hat \Theta_{j}^B\rangle \leq  \langle \hat N^A\hat N^B\rangle
$, which is tighter than (\ref{WARUJ}).
By combining (\ref{WARUJSTARYauto}) with (\ref{RELOLD}) we get 
\begin{equation}
\begin{multlined}
\label{OLDANDNAJS}
\sum_{j=1}^3 \langle (\hat \Theta_{j}^A 
+\hat \Theta_{j}^B )^2\rangle_{sep}  \\ \geq 
2(\langle \hat 
N^{A} \rangle  + \langle \hat N^{B}\rangle)_{sep}  + 
\langle (\hat N^{A}  - \hat 
N^{B})^2\rangle_{sep}.
\end{multlined}
\end{equation}
The new EPR-like necessary condition for separability differs from the one of Simon and Bouwmeester by the second term on the RHS of (\ref{OLDANDNAJS}). As the term is always non-negative, this is a stronger condition.
For the standard quantum optical model of inefficient detection (see the main text, or Sec.~\ref{LOSSES}) the new condition holds for any efficiency.
Note, that (\ref{RELOLD}) does not contribute anything 
to  the relation (\ref{OLDANDNAJS}),  because it is an operator identity. That is, the condition (\ref{OLDANDNAJS}) reduces to (\ref{WARUJSTARYauto}).

For normalized Stokes parameters the 
EPR-like separability condition, which is an analog of (\ref{OLDANDWRONG}), reads
\begin{equation} 
\begin{multlined}
\label{NEWANDWRONG}
\sum_{j=1}^3 \left \langle \left ({\hat S}_{j}^A 
+{\hat S}_{j}^B \right)^2\right \rangle_{sep}  \\ \geq 
\left \langle \hat \Pi^A\frac{2}{\hat N^A}\hat \Pi^A 
+\hat \Pi^B\frac{2}{\hat N^B}\hat \Pi^B \right \rangle_{sep}.
\end{multlined}
\end{equation} 
For a derivation, see \cite{zuko} (and see also \cite{MULTIPORT, MULTIPORT2} for its generalizations to $d$ modes). 
Entanglement detection with (\ref{NEWANDWRONG}) also depends on the detector efficiency, but for the considered bright squeezed vacuum state the threshold efficiency $\eta_{crit}$ decreases with growing $\Gamma$. The $\eta_{crit}$ is lower than $1/3$ for any finite $\Gamma$.

If one uses the Cauchy-like inequality~(\ref{WARUJNOWYauto}) and the identity  $\sum_{i=1}^3\hat{S}_i^2= \hat{\Pi} + \hat{\Pi} \frac{2}{\hat{N}}\hat{\Pi}$ (see \cite{zuko}), then the following tighter EPR-like separability condition emerges
\begin{eqnarray}
\label{NEWANDNAJS}
\sum_{j=1}^3 \left \langle \left ({\hat S}_{j}^A 
+{\hat S}_{j}^B \right)^2\right \rangle_{sep} \geq
\langle \hat \Pi^A\frac{2}{\hat N^A}\hat \Pi^A\rangle_{sep} \nonumber   \\ +  
 \langle \hat \Pi^B\frac{2}{\hat N^B}  \hat \Pi^B \rangle_{sep}
+ \langle (\hat \Pi^A   
-\hat \Pi^B)^2\rangle_{sep}.
\end{eqnarray}
It  is equivalent with the much 
simpler linear condition  (\ref{WARUJNOWYauto}).
The condition presented here has much more resistant to losses that the one derived in \cite{zuko}, and generalized in  \cite{MULTIPORT2},  here formula (\ref{NEWANDWRONG}).

\section{Resistance with respect to  losses}
\label{LOSSES}
%%%%%%%%%%%%%%%%%
Here we derive the dependence on a detector efficiency 
of  average values of entanglement indicators for optical fields 
$\hat{\mathcal{W}}_{\Theta}$ and $\hat{\mathcal{W}}_{S}$.
Our reasoning can be  extended  
to an arbitrary number of quantum optical modes and multi-party cases.

The loss model (an ideal detector and a beamsplitter of 
transmission amplitude $\sqrt{\eta}$ in front of it) is described by  a beamsplitter transformation for the creation operators, see e.g. \cite{KLYSHKO}, which  reads
\begin{equation}
\begin{multlined}
\label{ETAKREACJA}
\hat a^{\dagger}_j(\eta) % = \hat{\mathcal{U}}(\eta)\hat a^{\dagger}_j\hat{\mathcal{U}}^\dagger (\eta) 
= \sqrt{\eta} \hat 
a^{\dagger}_j+ \sqrt{1-\eta} 
\hat c^{\dagger}_j, 
\end{multlined}
\end{equation}
where $\hat{a}^{\dagger}_j$ refers to the  detection channel  in  $j$-th mode and
$\hat{c}^{\dagger}_j$ refers to the loss channel linked with the mode.

First, we shall analyze the problem  for standard Stokes operators. Let $\ket{\psi^{AB}}$ be a pure state of the modes,
before the photon losses. The unitary transformation $\hat{\mathcal{U}}(\eta)$ describing losses in all channels leads to $\hat{\mathcal U}(\eta)|\psi^{AB}\rangle = |\psi^{AB}(\eta)\rangle$, and we have
\begin{equation}
\expval{\mathcal{\hat W}_{{\Theta}}}{\psi^{AB}(\eta)} 
= \expval{\mathcal{\hat W}_{{\Theta}}(\eta)}{\psi^{AB}},
\end{equation}
where
$\mathcal{\hat{W}}_{{\Theta}}(\eta) = \hat{\mathcal{U}}^{\dagger}(\eta)\mathcal{\hat{W}}_{\Theta}\hat{\mathcal{U}}(\eta)$.
A transformed photon number operator $\hat n_j (\eta) = \hat a_j^{\dagger}(\eta)\hat a_j(\eta)$   reads
\begin{eqnarray}
\label{nOPER}
\hat n_{j}(\eta) &=&  (\sqrt{\eta} \hat 
a_{j}^{\dagger}+ \sqrt{1-\eta} 
\hat c_{j}^{\dagger}) (\sqrt{\eta} \hat 
a_j + \sqrt{1-\eta} \hat c_{j}) \nonumber \\
&=& \eta \hat n_{j} + \sqrt{\eta(1-\eta)}(\hat c_{j}^{\dagger} \hat a_{j} +
\hat a_{j}^{\dagger} \hat c_{j})
+(1-\eta) \hat c_{j}^{\dagger} \hat c_{j}. \nonumber \\
\end{eqnarray}
Notice that as the original state $\ket{\psi^{AB}}$ does not contain photons in the 
loss channels, thus in $ \expval{\hat n_{i}^A(\eta^{A})\hat n_{j}^B(\eta^{B})}{\psi^{AB}}
$ only the first term of the second line of~(\ref{nOPER}) survives. For the transmission amplitudes $\eta^{A}$ and $\eta^{B}$ of beams $A$ and $B$, we have
\begin{equation}
\expval{\hat n_{i}^A(\eta^A)\hat n_{j}^B(\eta^B)}{\psi^{AB}}  = \eta^A \eta^B
\expval{\hat n_{i}^A\hat n_{j}^B}{\psi^{AB}}. 
\end{equation}
From this we get the dependence of correlations of Stokes operators on detection efficiency in the form of $\langle\hat{ \Theta}_i^A(\eta^A) \hat{ \Theta}_j^B(\eta^B)\rangle=\eta^A\eta^B\langle\hat{ \Theta}_i^A \hat{ \Theta}_j^B\rangle$.

For normalized Stokes operators, the reasoning is  as follows.
For Fock states $|F\rangle=| n_{A_i}, n_{A_{i_\perp}}, m_{B_i}, m_{B_{i_\perp}}\rangle$, it is enough to consider only the average value of $\hat{S}^A_3$ for state $|F_A\rangle=|n_{A_H}, m_{A_{V}}\rangle$, which we shall denote for simplicity as $|n, m\rangle$. Obviously for such a state the intensity rate at the detector measuring output $H$, with the detection efficiency $\eta$ for each of the detectors in the station, reads
\begin{eqnarray*}
r_1(\eta)&=&\langle (n,m)_\eta| \hat{\Pi}_A \frac{\hat{n}_H}{\hat{n}_H+\hat{n}_V} \hat{\Pi}_A|(n,m)_\eta\rangle \nonumber \\
&=&\sum_{k=1}^{n}\sum_{l=0}^{m}{{n}\choose{k}}{{m}\choose{l}}\frac{k}{k+l}\eta^{k+l}(1-\eta)^{n+m-k-l}.
\end{eqnarray*}
First we notice that $k{{n}\choose{k}}={n} {{n-1}\choose{k-1}}$, and rewrite the first summation as from $k=0$ to $k=n-1$. 
Next, let us consider a function $f(\gamma, \eta)$ of the form 
\begin{eqnarray}
f(\gamma, \eta) 
&=&n\sum_{k=0}^{n-1}\sum_{l=0}^{m}{{n-1}\choose{k}}{{m}\choose{l}}\frac{1}{k+1+l} \nonumber \\
&&\times \gamma^{k+1+l}(1-\eta)^{n-1+m-k-l},
\end{eqnarray}
which for $\gamma=\eta$ gives $r_1(\eta)$. Its derivative with respect to $\gamma$
reads
\begin{eqnarray}
&&\frac{d}{d\gamma}f(\gamma, \eta)\nonumber \\
&=&n\sum_{k=0}^{n-1}\sum_{l=0}^{m}{{n-1}\choose{k}}{{m}\choose{l}}\gamma^{k+l}(1-\eta)^{n-1+m-k-l}\nonumber\\
&=&n(\gamma+1-\eta)^{n+m-1}.
\end{eqnarray}
This upon integration with respect to $\gamma$, with the initial condition $f(\gamma=0,\eta)=0$, gives for $\gamma=\eta$
the required result:
\begin{equation}
\label{RES}
r_1(\eta)=\frac{n}{n+m}\big(1-(1-\eta)^{n+m}\big).
\end{equation}

It is easy to see that this result has a straightforward generalization 
to the case of more than two local detectors (e.g., see Fig. 1 in the main text).
To calculate the dependence on $\eta$ of the rate at detector $i$, 
when we have altogether $d$ detectors at the station, we simply replace in the above formulas $\hat{n}_H$ by 
$\hat{n}_i$ and $\hat{n}_V$ by $\sum_{j\neq i} \hat{n}_j$, to get $ r_i(\eta)=\frac{n}{n_{tot}}(1-(1-\eta)^{n_{tot}})$, where $n$ is the number of photons in a Fock state in mode $i$
and $n_{tot}$ is the total number of photons.

Note that for four-mode bright squeezed vacuum state (\ref{BOH_PROJ_0}) our entanglement condition for normalized Stokes parameters (\ref{WARUJNOWYauto})  is fully resilient with respect to losses of the kind described above. This is due to the fact that   squeezed vacuum is a superposition entangled states (\ref{PSINONOISE}), and each of them violates the separability criterion. As the Stokes operators do not change overall photon numbers on each of the sides of the experiments which we consider here, and states $\ket{\psi^{n}_{-}}$ contain $n$ photons in both beams $A$ and $B$, an inefficient detection in the case of $\ket{\psi^{n}_{-}}$ introduces the same reduction factor on both sides of condition  (\ref{WARUJNOWYauto}). The violation of it holds for whatever value of $\eta$. The expectation values for the full squeezed state are simply weighted sum of expectation values for its components $|\psi^n_-\rangle$.  The same can be shown for all other squeezed states, and linear separability conditions considered here, including the cases of $d>2$.

%paprameters so introducing  losses (\ref{RES}) does not  influence violation of (\ref{WARUJNOWYauto}).}

%SECOND READING UP TO THIS POINT.

%%%%%%%%%%%%%%%%%
\section{Entanglement experiments involving multiport beamsplitters:  
homomorphism of single qudit observables and field operators}
\label{apx:qudit_density}
%%%%%%%%%%%%%%%%%
{\it Proof of relation (11) of the main text for qu$d$it states.}---We consider a  set of unitary qudit  observables of the following form in the main text  
\begin{eqnarray}
\label{SMALLq}
\hat{q}_k(m)= \sum_{j=1}^d\omega^{jk}|j(m)\rangle \langle j(m)|,
\end{eqnarray}
where $k=0,1,..., d-1$ and $\omega=\exp(2\pi i/d)$, and $\hat{U}(m) | j\rangle = | j(m)\rangle$ is a unitary transformation of a computational basis  ($m=0$) to a vector of a different unbiased  basis $m$. We assume that the bases $m\neq m'$ are all mutually unbiased, and consider only dimensions in which we have $d+1$ mutually unbiased bases. 
We show that the operators $\hat{q}_k(m)/\sqrt{d}$ with $k = 1,...,d-1$ 
and $m=0,...,d$, and $\hat{q}_0(0) = \openone$ form an orthonormal basis in the 
Hilbert-Schmidt space of  (all) $d\times d$  matrices.

The orthonormality  of the operators can be established as follows.
We are to prove that
\begin{eqnarray}
\frac{1}{d} \Tr 
\hat{q}_k^{\dagger}(m)\hat{q}_{k'}(m') = 
\delta_{mm'}\delta_{kk'}.
\label{ORTHO}
\end{eqnarray}

\begin{itemize}
\item For $k'=0$, this is trivial because all $k\neq 0$ 
operators are traceless (as $\sum_{j=1}^{d}\omega^{jk}=d\delta_{k0}$).
\item For $m\neq m'$, with $k\neq 0$ and $k'\neq 0$, one has
\begin{eqnarray}
&&\frac{1}{d}\sum_{l,j,j'} \omega^{-jk+j' k'} \langle l(m) |j(m)\rangle\langle j(m)| j'(m')\rangle\langle j'(m')|l(m)\rangle \nonumber \\
&=& \frac{1}{d} \sum_{l,j'}\omega^{-lk +j'k'}\langle l(m)| j'(m')\rangle\langle j'(m')|l(m)\rangle \nonumber \\ 
&=&\frac{1}{d^2}\sum_{l,j'} \omega^{-lk +j'k'} =0,
\label{EQ:ORTHO2}
\end{eqnarray}
where we use the fact that for mutually unbiased bases $|\langle j'(m')|j(m)\rangle|^2=1/d$. 
\item For $m=m'$, in the second line of 
(\ref{EQ:ORTHO2}) we have  \\ $\langle 
j'(m')|l(m)\rangle 
=\delta_{lj'}$, and we get in the last line
$\frac{1}{d}\sum_l\omega^{l(k-k')}= \delta_{kk'}.$
\end{itemize}
As we have $(d-1)(d+1)+1=d^2$ such orthonormal operators, the basis is complete. QED.

{\it Remarks on the homomorphism.}---We shall now show that for any pure 
state of a $d$-mode optical field $\ket{\psi}$, 
one can always find a $d \times d$ 
one qudit density matrix $\mathfrak{M}$ for which  the following holds
\begin{equation}
\frac{\expval{\hat Q_{k}(m)}{\psi}}{\expval{\hat \Pi}{\psi}} =  
\Tr \hat{q}_{k}(m)\mathfrak{M},
\label{eq:qudit_homo_rate}
\end{equation}
where $ \hat{Q}_{k}(m) $ is defined by (15) in the main text.
%This proof will equally be a confirmation of the relation (\ref{HELPREL}) 
%that is a direct analogue of(\ref{SINGLE2}) for optical field.
For the expectation value, which reads
\begin{eqnarray}\label{Q-OBS}
\bra{\psi} \hat{Q}_{k}(m)\ket{\psi}
= \bra{\psi} \sum_{j=1}^{d}  
\hat \Pi \frac{a_{j}^{\dagger} (m) 
a_{j}(m) }{\hat{N}}  \hat \Pi  \, \omega^{jk} 
\ket{\psi},
\end{eqnarray}
we introduce a set of states
\begin{equation}
\ket{\phi_{j}(m)} = \ani{j}(m) 
\frac{1}{\sqrt{\hat{ N}}}  \hat \Pi \ket{\psi},
\label{EQ:STATE}
\end{equation}
which for $m=0$ gives
\begin{equation}
\ket{\phi_{j}(0)} = \ani{j} \frac{1}{\sqrt{\hat{N}}}  
\hat \Pi \ket{\psi}.
\end{equation}
Then, one can transform (\ref{Q-OBS}) into
\begin{equation}
\avg{\hat{Q}_{k}(m)} = \sum_{j=1}^d 
\inner{\phi_{j}(m)}{\phi_{j}(m)} \omega^{jk}.
\label{EQ:EXP_obs}
\end{equation}

As it was mentioned in the main text, the unitary transformation of the creation operators between input and output beams is $\hat a_{l}^{\dagger} (m) = \sum_{r}  {U}_{l 
r} (m) \hat a_{r}^{\dagger}$,
where $\hat a^{\dagger}_r =\hat a^{\dagger}_r(m=0)$ is a reference operator and $ U(m=0) = \mathbb{1}$. Thanks to this the state~(\ref{EQ:STATE}) can be put as
\begin{eqnarray}
\ket{\phi_{j}(m)} &=& \sum_{s=1}^d U_{js}^{*} (m) \ani{s} 
\frac{1}{\sqrt{\hat{N}}} \hat \Pi \ket{\psi} \nonumber \\
&=& \sum_{s} U_{js}^{*} (m) \ket{\phi_{s}(0)}.
\end{eqnarray}
Therefore, (\ref{EQ:EXP_obs}) can be put as
\begin{eqnarray*}
\avg{\hat{Q}_{k} (m)} &=& \sum_{j,s,r=1}^d
\omega^{jk} \bra{\phi_{r}(0)} {U}_{jr} (m) 
U_{js}^* (m) \ket{\phi_{s}(0)}.
\end{eqnarray*}
Let us introduce  a matrix, denoted by $M$, whose elements are $M_{sr} = \inner{\phi_{r}(0)}{\phi_{s}(0)}$. Then 
\begin{eqnarray}
 \sum_{r,s=1}^d \bra{\phi_{r}(0)} 
U_{jr} (m) U^*_{js} (m) \ket{\phi_{s}(0)}
\label{MATRIXORI}
\end{eqnarray}
becomes
\begin{eqnarray}
 \sum_{r,s=1}^d  {U}_{jr} (m) 
M_{sr} U_{js}^* (m)  
=\left[ U (m) M^T U^{\dagger} (m) \right]_{jj}.
\end{eqnarray}
Finally we arrive at
\begin{eqnarray}
\avg{\hat{Q}_{k}(m)} =\sum_{j} \omega^{jk} \left[ U (m) M^T U^{\dagger} (m) \right]_{jj},
\end{eqnarray}
where $M$ is a (positive 
definite) Gramian  matrix. 
 Its trace 
is given by $\Tr M = \langle \hat \Pi \rangle  \leq 1$. 
We can normalize it to get  $\mathfrak{M} =M/\langle\hat\Pi\rangle$, which is  an admissible qudit density matrix.

Let us now turn back to qudits, and analyze the structure an expectation of the unitary observable~(\ref{SMALLq}). 
First, consider a pure state $\ket{\xi}$. The expectation value reads 
\begin{eqnarray}
\bra{\xi} \hat{q}_{k}(m) \ket{\xi}&=& 
\sum_{j,r,s} \omega^{jk} U_{js}(m) {U}^{*}_{jr} (m)\inner{r}{\xi} \inner{\xi}{s} \nonumber \\
&=&\sum_{j,r,s} \omega^{jk}  U_{js} (m) M_{rs}^{\xi} U^{*}_{jr} (m) \nonumber \\
&=&\sum_{j} \omega^{jk} \left[ U(m) M^{\xi T} U^{\dagger} (m) \right]_{jj},
\label{eq:exp_chi}
\end{eqnarray}
where we use $\ket{j (m)} = \sum_{r} U_{jr} (m) \ket{r}$ and introduce a density matrix $M^{\xi}$ for the state $\ket{\xi}$ of elements  $M_{rs}^{\xi} = \inner{r}{\xi}\inner{\xi}{s}$.
If we replace $\ket{\xi}$ by a density matrix given by ${\varrho} = \sum_{\lambda} p_\lambda \ket{\xi_\lambda} \bra{\xi_\lambda}$, then the expectation~(\ref{eq:exp_chi}) becomes
\begin{eqnarray}
\Tr\varrho \hat{q}_{k}(m) 
&=& \sum_{\lambda} p_\lambda \bra{\xi_\lambda} \hat{q}_{k}(m) 
\ket{\xi_\lambda} \nonumber \\
&=&\sum_{j} \omega^{jk} \left[ U(m) M^{\varrho T} U^{\dagger} (m) \right]_{jj},
%&=& \sum_{l,j} p_l \omega^{jk}  \left[ \overline{U}(m) M^{\chi_l} U^{T} (m)  \right]_{jj}.
\end{eqnarray}
where  matrix $M^{\varrho}$ has  elements given by $M^{\varrho}_{rs}=\sum_{\lambda} p_\lambda \inner{r}{\xi_\lambda} \inner{\xi_\lambda}{s}$.
Therefore, (\ref{eq:qudit_homo_rate}) holds. 
Obviously, such reasoning can be generalized to the case of (mixed) states describing correlated beams $A$ and $B$, in the way it is done in the main text.

For intensity-based observables, we have a similar relation 
\begin{equation}
\frac{ \expval{{\hat \chi}_{k}(m)}{\psi}} 
{\expval{\hat  N}{\psi} }  =  \Tr 
\hat{q}_{k}(m)\mathfrak{N},
\end{equation}
where $\mathfrak{N}$ is a possible two-qudit density matrix. Note that in general $\mathfrak{M} \neq 
\mathfrak{N}$.

%{\bf END OF WORK so far.}

\section{Noise resistance of 
Cauchy-Schwartz-like separability condition for Bright Squeezed Vacuum}
\label{NOISE}

Observables based on rates can in some cases allow a more noise resistant
entanglement detection than the ones based directly on intensities.

{\it Distortion noise.}---We take as our working example a  $d\times d$ mode bright squeezed vacuum
in the  presence of a specific type of noise, which can be treated 
as distortion of the state, 
which lowers the correlations between the beams.

\subsection{$2\times2$ mode bright squeezed vacuum plus noise}

We build our noise model in following steps. Let us introduce four squeezed vacuum states which are related with the Bell state basis for two qubits. 
To make our notation concise let us denote by $k=0$ the polarization $H$ and by $k=1$ polarization $V$, and let us define that the index values follow modulo 2 algebra.
Then one can write down the following

\begin{equation}
\label{BSV-psi}
 \ket{\Xi(m,l)} =\frac{1}{\cosh^2 \Gamma} \sum_{n=0}^{\infty} \frac{\tanh^n{\Gamma}}{n!}\left(\sum_k (-1)^{km}a^\dagger_kb^\dagger_{k+l}\right)^n|\Omega\rangle
\end{equation}
and define squeezed vacua related with the Bell states as  $\ket{\Xi(0,0)}  = \ket{\Phi^+}$,  $\ket{\Xi(0,1)}=\ket{\Psi^+}$, $\ket{\Xi(1,0)}=\ket{\Phi^-}$, and $\ket{\Xi(1,1)}=\ket{\Psi^-}$. This notation may look too dense here, but it will help us further on.
Our noise model, which is an analog of the ``white noise" in the case of two qubits, can be defined as
\begin{equation} 
\label{NOISE-1}
 \varrho^{noise} = \frac{1}{4}(\ketbra{\Psi^-} + \ketbra{\Psi^+} + \ketbra{\Phi^-} + \ketbra{\Phi^+}).
\end{equation}

The following properties of the noise are essential.
For each $i$ and $j$,
\begin{equation} 
\Tr 
 \hat S_i^{A}
\hat S_j^{B}\varrho^{noise} =0.
\end{equation}
That is the noise itself such that it leads to vanishing correlations between components of the Stokes parameters. This is easy to see when one recalls the local unitary transformations, say on side $A$, (replaced here by mode transformations) which link the three other two-qubit Bell states with the singlet. Simply they are equivalent to $\pi$ rotations of Bloch sphere of side $A$ with respect to axes $z$, $x$, and $y$.   
The second property is 
\begin{equation}
\expval{\hat \Pi^{A}\hat \Pi^B}{\Psi^-} 
= \Tr \hat \Pi^{A} 
\hat \Pi^{B} \varrho^{noise}.
\end{equation}

{\it For normalized Stokes operators.}---Let us start with the analysis of noise in terms 
of the rate observables. 
Let $v$ be the 
\textit{visibility}, which determines the following 
noisy state:
\begin{equation}
\varrho^{AB} = v\ketbra{\Psi^-} + (1-v)\varrho^{noise}, 
\end{equation}
where $0\leq v \leq1$.
%at the moment of the 
We have to find the threshold $v$ above which our separability condition
$\sum_{i=1}^3|\langle \hat S_i^A\hat S_i^B\rangle|_{sep} 
\leq \langle \hat\Pi^A\hat\Pi^B\rangle_{sep}$ 
fails to hold. It happens when 
\begin{equation} 
\label{MULTINOISE}
v\sum_{i=1}^{3} 
\left| \expval{{\hat S}_i^{A} { \hat S}_i^{B}}{\Psi^-}\right| >
\expval{\hat \Pi^{A}\hat \Pi^B}{\Psi^-}.
\end{equation}
This will be our measure of the resilience with respect to the noise.
 
Applying the technical  facts that  for $\ket{\Psi^-}$ one has $\expval{{\hat S}_i^{A}
{ \hat S}_i^{B}}{\Psi^-}=-\expval{({\hat S}_i^{A})^2}{\Psi^-}$ and $ \expval{ 
	\hat \Pi^{A}\hat \Pi^B 
}{\Psi^-}  = \expval{ 
\hat \Pi^{A} 
}{\Psi^-}$,
one gets
$$\sum_{i = 1}^3 
\left| \expval{\hat S_{{i}}^{A}
	\hat S_{{i}}^{B}}{\Psi^-} \right| = \expval{\hat \Pi^A +\hat \Pi^A
	\frac{2}{\hat N^A}\hat \Pi^A}{\Psi^-} $$ and
the condition for detection of entanglement  reads
\begin{equation}
v
 \expval{\hat \Pi^A + \hat \Pi^A
\frac{2}{\hat N^A}\hat \Pi^A}{\Psi^-}   >  \expval{ 
\hat\Pi^{A} 
}{\Psi^-}. 
\end{equation} 
The threshold visibility $v_{crit}$ is given by 
\begin{equation}
\label{NOISEcond}
v_{crit}   =  \frac{\expval{\hat \Pi^A}{\Psi^-}}{  \expval{\hat \Pi^A  
+ \hat{\Pi}^A\frac{2}{\hat N^A}\hat \Pi^A}{\Psi^-}}.
\end{equation}
The respective terms of  (\ref{NOISEcond}) are given by 
\begin{equation}
\label{NONVAC}
\expval{\hat \Pi^A}{\Psi^-} = 1 -\frac{1}{\cosh^{4}\Gamma} 
=  1- \sech^{4}\Gamma
\end{equation} 
that follows from the definition of $\langle \hat{\Pi}^A\rangle $ and  
\begin{eqnarray}
&& \expval{ \hat{\Pi}^A\frac{1}{\hat N^A} \hat{\Pi}^A}{\Psi^-} \nonumber \\
&&= \frac{2 \tanh^2\Gamma }{\cosh^{4}\Gamma}   
{}_3F_{2}(1,1,3;2,2;\tanh^2 \Gamma),
\end{eqnarray}
where ${}_3F_2(1,1,3;2,2;\tanh^2 \Gamma)$ is generalized hypergeometric function.
%of $_pF_q(a_1,...,a_p;b_1,...,b_q;x)$.

%EDITING UP TO THIS POINT.

{\it For standard Stokes operators.}---Following the same reasoning  for 
observables based rates 
the threshold visibility $v_{crit}^{old}$ for observables 
based on intensities 
is given by
\begin{equation}
\label{VIZold}
v_{crit}^{old} = \frac{\expval{(\hat N^{A})^{2}}{\Psi^-}}{ 
\expval{\hat N^A(\hat N^A + 2)}{\Psi^-}
}. 
\end{equation}
We have 
\begin{equation}
 \expval{\hat N^A}{\Psi^-} = 
 2\sinh^2 \Gamma
\end{equation}
and  
\begin{eqnarray}
%\begin{multlined} 
\expval{(\hat{N}^{A})^{2}}{\Psi^-} &=& \frac{2 \tanh^2 \Gamma}{\cosh^4 \Gamma} \frac{2\tanh^2 \Gamma+1}{(1-2\tanh^2 \Gamma)^4} \nonumber \\
&=& \sinh^2 \Gamma ( 3 \cosh 2\Gamma -1 ).
\label{NEWTERM}
%\end{multlined}
\end{eqnarray}

\begin{figure}[t]
 \includegraphics[scale= 0.8]{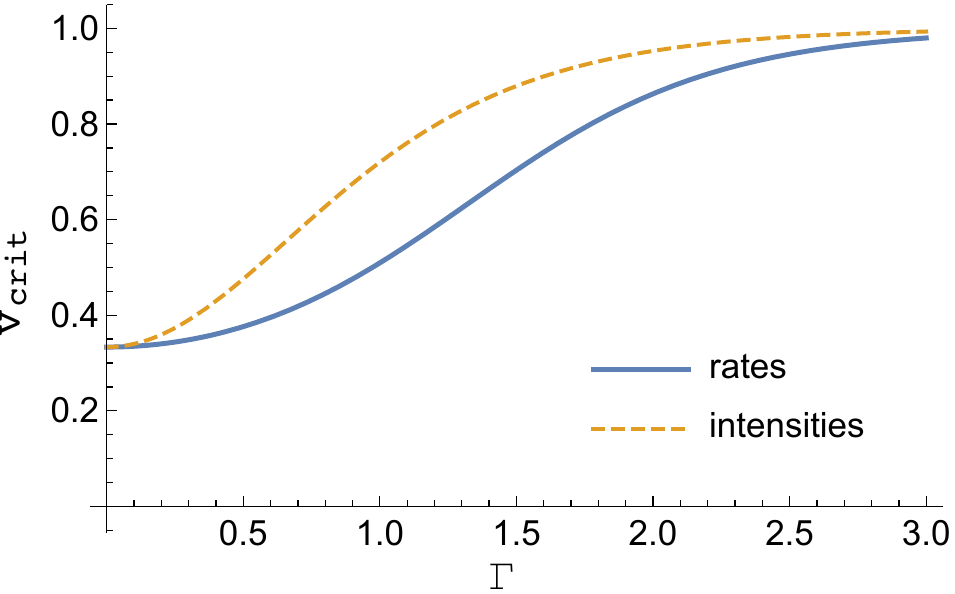}
 \caption{Comparison of critical visibilities
to detect entanglement via the Cauchy-like condition, for four-mode squeezed vacuum $\ket{\Psi^-}$ mixed with ``white" noise in (\ref{NOISE-1}).
The upper curve is for standard Stokes parameters, and the lower for normalized ones. The latter one turns out to lead to a higher noise resistance. Note that the Cauchy-like condition is equivalent in the case of  $\ket{\Psi^-}$, and the mixture of $\ket{\Psi^-}$ with the model noise, with the necessary and sufficient conditions to detect entanglement via measurement of Stokes parameters. Therefore, this graph shows the critical visibilities also for this case. One cannot do better.  Obviously the graphs for $\ket{\Psi^+}$, $\ket{\Phi^+}$ and $\ket{\Phi^-}$ are identical. The asymptotic limit $v_{crit}=1/3$, for $\Gamma \rightarrow 0$, is concurrent with the white noise threshold for a two-qubit singlet.}
\label{fig:viz_viZ}
\end{figure}

The form of  (\ref{NEWTERM}) was
obtained as follows. Let us put $x =\tanh^2{\Gamma}$, and $c=\cosh^4{\Gamma}$. We have
\begin{multline}
\langle  (\hat{N}^{A})^{2}\rangle  =
\frac{1}{c} \sum_{n=0}^{\infty}x^n(n+1)n^2 =
\frac{x}{c}\frac{d^2}{d x^2} \bigg(\sum_{n=0}^{\infty}nx^{n+1}
\bigg) \\=  
%\frac{x}{\cosh^4(\Gamma)}\frac{d}{d^2x}\bigg( x^2 \sum_{i=0}^{\infty} nx^{n-1} \bigg) 
%= \frac{x}{\cosh^4(\Gamma)}\frac{d}{d^2x}\bigg(x^2\sum_{i=1}^{\infty}\frac{d}{dx}x^n \bigg) \\=
\frac{x}{c}\frac{d^2}{dx^2}\bigg(x^2\frac{d}{dx}\sum_{n=0}^{\infty}x^n \bigg) =
\frac{x}{c}\frac{d^2}{dx^2}\bigg(x^2\frac{d}{dx}\bigg(\frac{1}{1-x}\bigg) \bigg) \\= \frac{2x(2x+1)}{c(1-x)^4}.
\end{multline}
Thus, the threshold visibility in function of the 
amplification gain $v_{crit}^{old}(\Gamma)$ for the ``macroscopic singlet" $\ket{\Psi^-}$
is
\begin{equation}
v_{crit}^{old}(\Gamma)= \frac{3\cosh{2\Gamma} -1}{3\cosh{2\Gamma} +3}.
\end{equation}
We compare the critical visibilities obtained with the two approaches (normalized vs. standard Stokes parameters) in Fig. \ref{fig:viz_viZ}.

%END of EDITING HERE

\subsection{Unitary observables for $d$-mode}

{\it Multimode bright squeezed vacuum.}---The bright squeezed vacuum is a state of light 
of 
undefined photon number which has, due to entanglement,  perfect EPR correlations 
of numbers of photons between specific modes reaching $A$ and $B$.  
Such an entanglement can be observed in multimode 
parametric down-conversion  emission. The
 interaction
Hamiltonian of the process, for a classical pump,  is essentially 
$\hat H =i \gamma  \sum_{j=0}^{d-1}\hat a_j^{\dagger}\hat b_j^{\dagger} + h.c.$
where $\gamma$ is the coupling constant proportional to a pump power. Thus, 
$d\times d$ mode (bright) squeezed vacuum state is given by
\begin{equation}
\label{MULTIVAC}
 \ket{\Psi_{BSV}^{d}} =\frac{1}{\cosh^d \Gamma} \sum_{n=0}^{\infty} 
 \sqrt{\frac{(n+d-1)!}{n!(d-1)!}} \tanh^n\Gamma \ket{\psi^{n}_d},
\end{equation}
where  $\Gamma  = \gamma t$ and $t$ is the interaction time, and 
\begin{eqnarray}\label{PSIN}
\ket{\psi^n_d}= \sqrt{\frac{n!(d-1)!}{(n+d-1)!}} \frac{1}{n!} \left(\sum_{j=0}^{d-1}\hat a_j^{\dagger}\hat 
b_j^{\dagger}\right)^n\ket{\Omega}.
\end{eqnarray}

{\it Noise model.}--- If we consider the unitary observables, our noise model can look as follows: we build our noise model in following similar steps as for the $d=2$ case. 
%Let us introduce four squeezed vacuum states which are related with the Bell state basis for two qudits. 
Let us now index $k$ stand for local modes $k=0,1,...,d-1$ and we shall  the modulo $d $  algebra for it.
Then one can write down the following
\begin{equation}
\label{BSV-psi_d}
 \ket{\Xi^d(m,l)} =\frac{1}{\cosh^d \Gamma} \sum_{n=0}^{\infty} \frac{\tanh^n{\Gamma}}{n!}\left(\sum_k \omega^{km}a^\dagger_kb^\dagger_{k+l}\right)^n|\Omega\rangle
\end{equation}
with $m$ and $l$ taking values $0,1,...,d-1$. Note that these squeezed $d$-mode vacua are analogs of the following Bell basis for a pair of qudits: 
$\frac{1}{\sqrt{d}}\sum_k \omega^{km}|k\rangle\otimes|k+l\rangle$.
Our noise model is defined as
\begin{equation} \label{NOISE-2}
\varrho^{noise} = \frac{1}{d^2} \sum_{m,l}  \ketbra{\Xi^d(m,l)}.
\end{equation}

The following properties of the noise are essential for us.
For each $i$ and $j$
\begin{equation} 
\label{NOISE-PROPERTY}
\Tr 
 \hat{Q}_i^{A}(m)
\hat Q_j^{B\dagger}(m')\varrho^{noise} =0,
\end{equation}
%That is the noise itself is such that it leads to vanishing correlations between components of the Stokes parameters. This is easy to see when one recalls the %local unitary transformations, say on side $A$, (replaced here by mode transformations) which link the three other two-qubit Bell states with singlet. Simply %they are equivalent to $\pi$ rotations of Bloch sphere of side $A$ with respect to axes $z$, $x$, and $y$.   
and the second property is 
\begin{equation}
\Tr (\hat \Pi^{A} \hat \Pi^{B} \varrho^{noise})=\expval{\hat \Pi^{A}\hat \Pi^B}{\Psi^d_{BSV}}.
\end{equation}
%and 
%\begin{equation}
%\Tr( \hat N^{A} \hat N^{B} \varrho^{noise})=\expval{\hat N^{A}\hat N^B}{\Psi^d_{BSV}} . 
%\end{equation}
We have the same relation for observables based on intensities.
%Note that as $d$ increases the results obtained via rate and 
%intensity approach become similar.
%is given on  
%
%\begin{figure}[t]
%\centering
%{\includegraphics[width=8cm]{ratio}}
%\caption{The ratio between the threshold visibilities
%for rate and intensity,
%$v_{crit}^{rate}/v_{crit}^{int}$ for $d=2$ as a function of $\Gamma$. Note that 
%the biggest difference between two approaches in favor of rate-observables is 
%for $\Gamma \approx 1$.}
%\label{fig:RATIO}
%\end{figure}

{\it Noise resistance.}---Applying  this model we get that
entanglement detection is possible with the Cauchy-like condition for observables based on rates, in the case of $\ket{\Psi^d_{BSV}}$ mixed with the noise, if the threshold 
visibility $v_{crit}$ fulfills
\begin{equation}
\label{VIZnewUNI}
v_{crit}   =  \frac{\expval{\hat 
		\Pi^A}{\Psi^d_{BSV}}}{ 
	 \expval{\hat \Pi^A  + \hat 
		 \Pi^A\frac{d}{\hat N^A}\hat \Pi^A}{\Psi^d_{BSV}}}.
\end{equation}
In case of observables based on intensities, we get
\begin{equation}
\label{VIZoldUNI}
v_{crit}^{old} = \frac{\expval{(\hat{N}^{A})^{2}}{\Psi^d_{BSV}}}{ 
	\expval{\hat N^A(\hat N^A + d)}{\Psi^d_{BSV}}
}.
\end{equation}

\subsubsection{$3\times 3$ mode bright squeezed vacuum}
In case of  observables based on rates, the respective terms in (\ref{VIZnewUNI}) are as follows:
\begin{eqnarray} 
&&\expval{\hat \Pi^A\frac{3}{\hat N^A}\hat \Pi^A}{\Psi^3_{BSV}} \nonumber  \\ 
&&= \frac{1}{\cosh^{6}\Gamma} 9 \tanh^2\Gamma 
_3F_2(1,1,4;2,2;\tanh^2
\Gamma)
\end{eqnarray}
and 
\begin{equation}
\expval{\hat \Pi^A}{\Psi^3_{BSV}}  
=  1- \sech^{6} \Gamma.
\end{equation}

\begin{figure}[h]
	\includegraphics[scale=0.8]{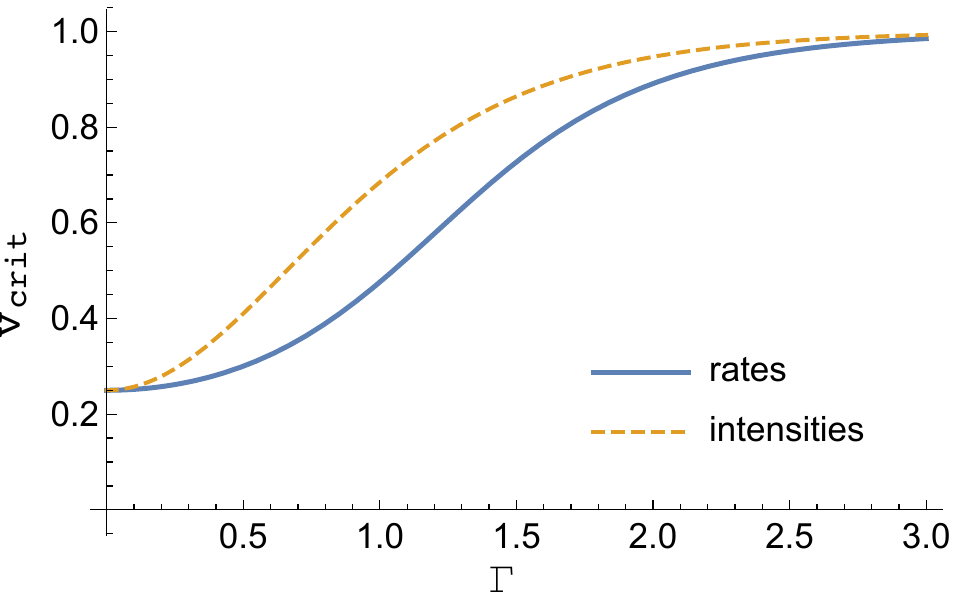}
\caption{Noise resistance for $\ket{\Psi^3_{BSV}}$. Note that for a very weak pumping the visibility approaches $1/4$.}
\label{FIG:suppFigS2}
\end{figure}

For  observables based on intensities in (\ref{VIZoldUNI}) we have
\begin{equation}
\expval{\hat N^A}{\Psi^3_{BSV}} = 
3\sinh^2 \Gamma
\end{equation}
 and 
 \begin{eqnarray}
 \label{expN2UNI}
 \expval{(\hat{N}^{A})^{2}}{\Psi^3_{BSV}}  
 &=&\frac{1}{\cosh^{6}\Gamma}
 \frac{3\tanh^2\Gamma(3\tanh^2\Gamma+1)}{(1-\tanh^2\Gamma)^5}  \nonumber \\
&=& 3\sinh^2\Gamma +  12\sinh^4\Gamma
 \end{eqnarray}
 The first equality of (\ref{expN2UNI}) can be obtained as (here, $x = \tanh^2\Gamma$ and $c =\cosh^6\Gamma$):
 \begin{eqnarray}
\langle (\hat {N}^{A})^{2}\rangle &=& \frac{1}{c} \sum_{n=0}^{\infty} \frac{(n+1)(n+2)}{2}n^2x^n \nonumber \\
&=& \frac{x}{2c}\frac{d^2}{dx^2} \bigg(\sum_{n=0}^{\infty} n(n+2)x^{n+1} \bigg) \nonumber \\
&=& \frac{x}{2c}\frac{d^2}{dx^2}\bigg(\frac{d}{dx}\bigg(\sum_{n=0}^{\infty} nx^{n+2}\bigg) \bigg) \nonumber \\
&=& \frac{x}{2c}\frac{d^3}{dx^3}\bigg(x^3 \frac{d}{dx}\sum_{n=0}^{\infty}x^n \bigg) \nonumber \\
&=& \frac{x}{2c}\frac{d^3}{dx^3}\bigg(x^3\frac{d}{dx}\bigg(\frac{1}{1-x}\bigg) \bigg) \nonumber \\
&=& \frac{1}{c}\frac{3x(3x+1)}{(1-x)^5}.
\end{eqnarray}
The threshold visibility in function of the 
amplification gain, $v_{crit}(\Gamma)$, for the macroscopic 
singlet $\ket{\Psi^3_{BSV}}$
is presented in Fig.~\ref{FIG:suppFigS2}.

%Nice!  So for $d$  if $\Gamma \to 0 $ $v_{crit} = \frac{1}{d +1}!$ 
%{\color{red}DO WE HAVE READY FORMULAS fOR a GENERAL $d$?}

\section{Derivation of some formulas used in Sec. \ref{NOISE}, and to obtain the general Cauchy-like separability condition}

\subsection{Formula 1} 
We shall show the following:
\begin{eqnarray}
\sum_{m=0}^{d} \sum_{k=1}^{d-1} |\hat{Q}_{k} (m)|^2
= (d-1) \left( \hat \Pi + \hat \Pi \frac{d}{\hat N}\hat \Pi \right).
\label{apx:abssquare}
\end{eqnarray}
%where $ |\hat{Q}_{k}^{X} (m)|^2=  \hat{Q}_{k}^{X } (m)\hat{Q}_{k}^{X\dagger } (m)$.
Note that this is a generalization of the identity $\sum_{i=1}^3\hat{S}_i^2= \hat{\Pi} + \hat{\Pi} \frac{2}{\hat{N}}\hat{\Pi}$.

The field operators involving the unbiased interferometers,  within the  approach with  rates (15) in the main text can be put as 
%(for simplicity, we omit $X$)
%\begin{eqnarray}
%\hat{Q}_{k} (m) = \sum_{j,l,l'=1}^{d} \hat{\Pi} \frac{U_{j l'}(m) \cre{l'} U^{*}_{j l}(m) \ani{l}}{\hat{N}}  \hat \Pi \,\, \omega^{jk}.
%\end{eqnarray}
%Thus,
\begin{eqnarray*}
\hat{Q}_{k} (m) = \sum_{l,l'=1}^{d} \left( \sum_{j=1}^{d} \omega^{jk} U_{j l'}(m)  U^{*}_{j l}(m)  \right) \hat{\Pi} 
\frac{\cre{l'} \ani{l}}{\hat{N}} \hat \Pi,
\end{eqnarray*}
and the formula for $\hat{Q}_{k}^{\dagger}$ is the Hermitian conjugate of the above.
The following relations
\begin{eqnarray}
\sum_{j=1}^d {U}_{j l'}(m) U^*_{j l}(m) \omega^{jk} = 
[ \hat{q}_{k} (m) ]_{l l'}
\end{eqnarray}
and
\begin{eqnarray}
\sum_{j=1}^d U^*_{j l'}(m) {U}_{j l}(m) \omega^{-jk} = 
[ {\hat{q}^{\dagger}_{k} (m)}]_{l l'}
\end{eqnarray}
lead to
\begin{eqnarray}
\hat{Q}_{k} (m) = \sum_{l,l'=1}^{d} [ \hat{q}_{k} (m) ]_{l l'} \hat{\Pi} \frac{\cre{l'} \ani{l}}{\hat{N}} \hat \Pi,
\end{eqnarray}
where $\hat{q}_k(m)$ are the qudit operators (\ref{SMALLq}).
Therefore, we have
%\begin{widetext}
\begin{eqnarray}
&&\sum_{m=0}^{d} \sum_{k=1}^{d-1}|\hat{Q}_{k} (m)|^2 = \hat \Pi \frac{1}{\hat{N}} \sum_{l l' n n'=1}^d \left( \sum_{m=0}^{d}\sum_{k=1}^{d-1}  \right. \nonumber \\
&&\times [ \hat{q}_{k} (m) ]_{l l'}  [{\hat{q}^{\dagger}}_{k} (m) ]_{n n'} \bigg) \cre{l'} \ani{l} \cre{n} \ani{n'}   \frac{1}{\hat{N}}\hat \Pi.
\end{eqnarray}
%\end{widetext}
As the operators $\frac{1}{\sqrt{d}}\hat{q}_k(m)$ and $\hat{q}_0(0)=\openone$ form  an orthonormal  basis in the Hilbert-Schmidt space of $d \times d$ matrix, we have 
$$
\delta_{l l'} \delta_{n n'} + 
\sum_{m=0}^{d} \sum_{k=1}^{d-1} [ \hat{q}_{k} (m)]_{l 
l'} 
[ {\hat{q}^{\dagger}}_{k} (m) ]_{n n'} = d
\delta_{ln}\delta_{l'n'}.
$$
All that, and $[a_{i}, a_{j}^{\dagger}] = \delta_{ij}$,  allow one to perform the following calculation:
%\begin{widetext}
\begin{eqnarray} \label{MULT}
&&\sum_{m=0}^{d} \sum_{k=1}^{d-1} |\hat{Q}_{k} (m)|^2 \nonumber \\
%&=& \hat \Pi \frac{1}{\hat{N}} \sum_{l l' n n'=1}^d \left( \sum_{m=0}^{d} \sum_{k=1}^{d-1}  [ \hat{q}_{k} (m) ]_{l l'}     [{\hat{q}^*}_{k} (m) ]_{n n'} \right) \cre{l'} \ani{l} \cre{n} \ani{n'}   \frac{1}{\hat{N}}\hat \Pi \nonumber \\
&=& \hat \Pi \frac{1}{\hat{N}} \sum_{l l' n n'=1}^d [ d\delta_{ln}\delta_{l'n'} - \delta_{ll'}\delta_{nn'}] 
\cre{l'} \ani{l} \cre{n} \ani{n'} \frac{1}{\hat{N}}\hat \Pi \nonumber \\
&=& \hat \Pi \frac{1}{\hat{N}} \left[ -\sum_{l  n}  \cre{l} \ani{l} \cre{n} \ani{n} + d\sum_{l l'} \cre{l'} \ani{l} \cre{l} \ani{l'} \right]  
\frac{1}{\hat{N}}\hat \Pi  \nonumber \\
%&=& \hat \Pi \frac{1}{\hat{N}} \left[ -\hat{N}^{2} + d \sum_{l l'=1}^d\cre{l'} \ani{l} ( \ani{l'} \cre{l} - \delta_{ll'}  )\right]  \frac{1}{\hat{N}}\hat \Pi \nonumber \\
&=& \hat \Pi \frac{1}{\hat{N}} \left[ -\hat{N}^2 + d \sum_{l l'} \cre{l'} \ani{l'} ( \cre{l} \ani{l} +1 ) - d \sum_{l} \cre{l} \ani{l}  \right]  
\frac{1}{\hat{N}}\hat \Pi  \nonumber \\
&=& \Pi \frac{1}{\hat{N}} \left[ -\hat{N}^2 + d\hat{N}^2 + d^2\hat{N} - d\hat{N}  \right]  \frac{1}{\hat{N}}\hat \Pi \nonumber \\
&=& \hat \Pi \left[ (d-1) + \frac{d(d-1)}{\hat{N}} \right]  \hat \Pi.
\end{eqnarray}
%\end{widetext}
 Thus, (\ref{apx:abssquare}) holds.

An analogue relation for the  observables involving intensities, which reads 
\begin{eqnarray}
\sum_{m=0}^{d} \sum_{k=1}^{d-1} | \hat{\chi}_{k} (m)|^2 = (d-1)\hat N(\hat N+ d),
\label{apx:abssquareint}
\end{eqnarray}
can be obtained by similar steps. It is a generalization of (\ref{RELOLD}).

\subsection{Formula 2}
\label{TECHNICAL_LEMA}
We here calculate the  expressions which enter 
of Cauchy-Schwartz-like  separability conditions 
based on rates~(14) and intensities~(16) in the main text 
for a $d\times d$ mode bright squeezed vacuum. Some of the formulas are also used in the discussion of noise resistance.

Let us consider first the condition~(16) in the main text: its LHS and RHS read
\begin{eqnarray}
\label{WARUJSINGLET}
\rm{LHS} &=& \sum_{m=0}^d \sum_{k=1}^{d-1} 
\left|\expval{\hat \chi_{{k}}^A(m)\hat  
\chi_{{k}}^{{B\dagger}}(m)}{\Psi^d_{BSV}} \right|, \nonumber \\
\rm{RHS} &=& (d-1)\expval{ (\hat N^{A})^{2}}{\Psi^d_{BSV}}.
\end{eqnarray}
 To get the formula for RHS we used
\begin{equation}
\expval{\hat N^A \hat N^B}{\Psi^d_{BSV}} = 
\expval{ (\hat N^{A})^{2}}{\Psi^d_{BSV}}.
\end{equation}
The action of  
$\hat \chi_k^B(m = 0)$ 
on an unnormalized 
$\ket{\psi^n_d}$ of (\ref{PSIN}),  which we put as  $\ket{\phi^n} = \left(\sum_{j=1}^d\hat 
a_j^{\dagger}\hat 
b_j^{\dagger}\right)^n\ket{\Omega}$,  is as follows
\begin{equation}
\label{ROBO}
 \hat \chi_k^B
\left(\sum_{j=1}^d\hat a_j^{\dagger}\hat 
b_j^{\dagger}\right)^n\ket{\Omega}
= \left(\sum_{j=1}^d\omega^{jk}
\hat b_j^{\dagger}\hat b_j\right)
\left(\sum_{j=1}^d\hat a_j^{\dagger}\hat 
b_j^{\dagger}\right)^n\ket{\Omega}.
\end{equation} 
%Let us remind that $[\hat b_{H}^{\dagger}, \hat 
%a_{H}^{\dagger}] = [\hat b_{H}, \hat a_{V}^{\dagger}] 
%=[\hat b_{H}, \hat a_{H}^{\dagger}]
% = [\hat b_{V}, \hat a_{H}^{\dagger}] =0$.
Let us denote as
$\hat X \equiv  \hat{\chi}_k^B =\sum_{j=1}^d \omega^{jk}\hat b_j^{\dagger}\hat b_j$ and $\hat Y \equiv
\sum_{j=1}^d\hat a_j^{\dagger}\hat b_j^{\dagger}$.
Then, we have
\begin{equation}
[\hat X, \hat Y] =  
 \left[\sum_{j=1}^d \omega^{jk}\hat b_j^{\dagger}\hat b_j,
 \sum_{j=1}^d\hat a_j^{\dagger}\hat b_j^{\dagger} \right] = 
\sum_{j=1}^d\omega^{jk}\hat a_j^{\dagger}\hat 
b_j^{\dagger}.
\end{equation} 
Next, we use the algebraic fact that if 
$[[\hat X,\hat Y],\hat Y] =0$, then the following holds $[\hat X,\hat 
Y^n] = n[\hat X, \hat Y]\hat Y^{n-1}$ and  
$ \hat X\hat Y^n = \hat Y^n\hat X + n[\hat X, \hat Y]\hat Y^{n-1}$. 
Applying this relation to (\ref{ROBO}) we get
\begin{eqnarray}
&&{\hat \chi}_k^B 
\left(\sum_{j=1}^d\hat a_j^{\dagger}\hat b_j^{\dagger}\right)^n\ket{\Omega} 
\nonumber
= \\ 
&&n \left(\sum_{j=1}^d \omega^{jk}\hat a_j^{\dagger}\hat b_j^{\dagger}\right) 
 \left(\sum_{j=1}^d \hat a_j^{\dagger} \hat b_j^{\dagger}\right)^{n-1}\ket{\Omega},
\label{apx:relation}
\end{eqnarray}
where we use $\sum_{j=1}^d \omega^{jk}\hat b_j^{\dagger}\hat b_j \ket{\Omega} =0$. 
We have the same relation if we replace $\hat \chi_k^B$ by  $\hat \chi_k^A$ in (\ref{apx:relation}), i.e.,
\begin{equation}
\label{RESULT}
{\hat \chi}_k^A \left(\sum_{j=1}^d
\hat a_j^{\dagger}\hat b_j^{\dagger}\right)^n\ket{\Omega} = 
{\hat \chi }_k^B \left(\sum_{j=1}^d \hat a_j^{\dagger}
\hat b_j^{\dagger}\right)^n\ket{\Omega}.
\end{equation} 
The identity (\ref{RESULT}) holds for all $m=0,1,\dots,d$.  In the case of $m\neq 0$ the formulas look the same if one employs creation and annihilation operators related with the interferometers $U(m)$ for $A$ and ${U}^\dagger (m)$ for $B$, and the fact that $\sum_{j=1}^d \hat a_j^{\dagger}
\hat b_j^{\dagger}= \sum_{j=1}^d \hat a_j^{\dagger}(m)
\hat b_j^{\dagger}(m)$, which is at the root of EPR correlations of the state.    All that, and  the identity (\ref{apx:abssquareint}),  lead to 
\begin{eqnarray}
\label{RESULT1}
\text{LHS} &=& \sum_{m=0}^d \sum_{k=1}^{d-1} \expval{|{\hat \chi}_{{k}}^A(m)|^2}{\Psi^d_{BSV}} \nonumber \\
&=&  (d-1)\expval{\hat N^A(\hat N^A+d)}{\Psi^d_{BSV}}.
\end{eqnarray}
Thus, we get
\begin{eqnarray}
\label{WARUNEK}
\expval{\hat N^A(\hat N^A+d)}{\Psi^d_{BSV}} 
> \expval{ ({\hat N^{{A}}})^2 }{\Psi^d_{BSV}}, \nonumber \\
\end{eqnarray}
for every $\Gamma$. 

A reasoning  following similar steps leads to a violation of  the Cauchy-Schwartz-like separability condition (14) in the main text for observables involving rates, as for the bright squeezed vacuum we have in this case:
\begin{eqnarray}
\label{TECHNICAL_NEW}
\expval{(\hat \Pi^A + \hat{\Pi}^A\frac{d}{\hat N^A}\hat \Pi^A)}{\Psi^d_{BSV}} >\expval{\hat \Pi^A }{\Psi^d_{BSV}}, \nonumber \\
\end{eqnarray}
where  $\hat \Pi^A = (\hat \Pi^A)^2$ was used.

\subsection{Property (\ref{NOISE-PROPERTY}) of the noise model}
%\widetext
The essential property of our noise is that for each $i$ and $j$ we get
\begin{equation} 
\label{COOLPROP}
\Tr 
\hat{\chi}_i^{A}(m)
\hat{\chi_j}^{B\dagger}(m')\varrho^{noise} =0,
\end{equation}
and we have the same relation for observables based on rates. We shall prove (\ref{COOLPROP}) for $d>2$. For simplicity we will use the intensity approach. The proof for the rate observables is similar.
 
For an arbitrary $d$ all  Bell-like maximally entangled states $ \ket{\Xi^d(k,l)}$ are linked by a unitary transformation that acts on one subsystem. The transformation is  as follows: 
\begin{eqnarray}
\label{TRANSSS}
 \hat{ \mathcal U}^{\dagger}(k,l)\hat b^{\dagger}_n\hat{\mathcal U}(k,l)  = \sum_{i=0}^{d-1}U(k,l)_{ni}\hat b^{\dagger}_i = \omega^{nk}\hat b^{\dagger}_{n+l}, %= \hat b^{\dagger}_n(m,l).
\end{eqnarray}
where $\hat b^{\dagger}_n$ stands for $k=0$ and $l=0$. Respectively, for annihilation operators we have:
$
 \hat{\mathcal U}^{\dagger}(k,l)\hat b_n\hat {\mathcal U}(k,l) = \sum_{i=0}^{d-1}\bar{U}(k,l)_{ni}\hat b_i =
 \omega^{-nk} \hat b_{n+l}.
$
Using transformation (\ref{TRANSSS}) we can present any $ \ket{\Xi^d(k,l)} $ as follows:  
\begin{eqnarray}
\label{BSV-psi_d2}
&&\ket{\Xi^d(k,l)} \nonumber \\
&=& \frac{1}{\cosh^d \Gamma} \sum_{n=0}^{\infty} \frac{\tanh^n{\Gamma}}{n!}\left(\sum_m a^\dagger_m \hat{\mathcal U^{\dagger}}(k,l)  b^\dagger_{m}\hat{\mathcal U}(k,l)\right)^n|\Omega\rangle \nonumber \\  
&=& \hat{\mathcal U^{\dagger}}(k,l) |\Psi^d_{BSV}\rangle.
\end{eqnarray}
Because this transformation is unitary we can replace the action of (\ref{TRANSSS}) on the state by its action on  the observables.
Thus, in order to prove (\ref{COOLPROP}) we shall show that  for any $i,j\neq 0$
\begin{equation}
\label{NOISE_1} \langle{\Psi_{BSV}^d}|
\sum_{k,l=0}^{d-1} \hat{\chi}_i^{A{}}(m)
\hat{\mathcal  U}(k,l)\hat{\chi}_j^{B\dagger}(m')\hat{\mathcal U}^{\dagger}(k,l)|{\Psi_{BSV}^d}\rangle  = 0.
\end{equation} 
It turns out that the above holds because of the following operator identity
\begin{equation}
\label{NOISE_MZ}
\sum_{k,l=0}^{d-1}
\hat{\mathcal{ U}}(k,l)\hat{\chi}_j^{B \dagger}(m')\hat{\mathcal{ U}}^{\dagger}(k,l)  = 0.
\end{equation}

The reverse of transformation 
(\ref{TRANSSS}) can be expressed in the following way:
\begin{equation}
 \hat{ \mathcal U}(k,l)\hat b^{\dagger}_n\hat{\mathcal U}^\dagger (k,l)  = \sum_{i=0}^{d-1}U^{-1}(k,l)_{ni}\hat b^{\dagger}_i = \omega^{-nk}\hat b^{\dagger}_{n-l}. %= \hat b^{\dagger}_n(m,l).
\end{equation}
Note that  $U^{-1}$  can  be decomposed as follows $
U^{-1}= Z^kX^l$, 
where $(Z^k)_{rn} = \delta_{rn}\omega^{-nk}$, $(X^l)_{ni} = \delta_{(n-l)i} $.  
Using the notation introduced above we get
%\begin{equation}
%\begin{multlined}
%\hat{\mathcal{U}}(k,l)\hat b^{\dagger}_r(m^{\prime})\hat{\mathcal{U}^{\dagger}}(k,l)
%=\hat{\mathcal{U}}(k,l)\hat{\mathcal{U}^{\dagger}}(m^{\prime})\hat b^{\dagger}_s
%\hat{\mathcal{U}}(m^{\prime})
%\hat{\mathcal{U}^{\dagger}}(k,l)
%\\ 
%=\hat{\mathcal{U}}(k,l)\sum_s^{d-1} U_{rs}(m^{\prime})\hat b^{\dagger}_r\hat{\mathcal{U}^{\dagger}}(k,l)
%= \sum_s^{d-1} U_{rs}(m^{\prime})\hat b^{\dagger}_s  U_{rs}(m^{\prime})(Z^kX^l)_{st}
%\hat b^{\dagger}_t.
%\end{multlined} 
%\end{equation}
%Thus, we get
\begin{eqnarray}
%\begin{multlined} 
&&\hat{\mathcal{U}}(k,l)\hat{\chi}^{B \dagger}_j \hat{\mathcal{U}^{\dagger}}(k,l) \nonumber \\  
&=& \hat{\mathcal{U}}(k,l) \sum_{r=0}^{d-1}\omega^{-rj}\hat{b}^{\dagger}_r(m^{\prime}) \hat{b}_r(m^{\prime})\hat{\mathcal U} ^{\dagger}(k,l) \nonumber \\ 
&=& \sum_{r=0}^{d-1}\omega^{rj}\sum_{s=0}^{d-1} U_{rs}(m^{\prime})\sum_{t=0}^{d-1}(Z^kX^l)_{st}\hat b^{\dagger}_t  \nonumber \\ &\times& \sum_{s^{\prime} =0}^{d-1} U^*_{rs^{\prime}}(m^{\prime})\sum_{t^{\prime} =0}^{d-1}(Z^kX^l)^*_{s^{\prime}t^{\prime}} \hat b_{t^{\prime}}.
\label{in_prog1}
%=\sum_{r=0}^{d-1}\omega^{rj} \sum_{t,t^{\prime} = 0}^{d-1}
%\sum_{s=0,s^{\prime}}^{d-1} U_{rs}(m^{\prime})(Z^kX^l)_{st}\bar U_{rs^{\prime}}(m^{\prime})(\bar Z^kX^l)_{s^{\prime}t^{\prime}}
% \hat b^{\dagger}_t
%\hat b_{t^{\prime}}
%\end{multlined}
\end{eqnarray}
We have 
\begin{eqnarray}
%\begin{multlined}
\label{in_prog2}        
& &\sum_{k,l=0}^{d-1}(Z^kX^l)_{st} (\bar Z^kX^l)_{s^{\prime}t^{\prime}}  \nonumber \\
&=& \sum_{k,l=0}^{d-1} \omega^{-sk}\delta_{(s-l)t} \omega^{s^{\prime}k} \delta_{(s^{\prime}-l) t^{\prime}}  \nonumber \\ 
&=& \sum_{l=0}^{d-1}\delta_{(s-l)t} \delta_{(s^{\prime}-l) t^{\prime}}\sum_{k=0}^{d-1} \omega^{-k(s-s^{\prime})} \nonumber \\
&=& \delta_{s,s^{\prime}} \delta_{t,t^{\prime}}.
%\end{multlined}
\end{eqnarray}
Combining (\ref{in_prog1}) and (\ref{in_prog2}) we get
\begin{eqnarray}
&&\sum_{k,l=0}^{d-1}\hat{\mathcal{U}}(k,l)\hat{\chi}^{B \dagger}_j \hat{\mathcal{U}^{\dagger}}(k,l) \nonumber \\ 
&=& \sum_{r=0}^{d-1}\omega^{-rj} \sum_{s,t=0}^{d-1} U_{r s}(m^{\prime})  U^*_{rs}(m^{\prime}) \hat b_{t}^{\dagger}\hat b_{t} \nonumber \\
&=& \sum_{r=0}^{d-1}\omega^{-rj}\hat N^B =0.
\end{eqnarray}
Thus the identity (\ref{NOISE_MZ}) holds. 

%{\bf Please check appendix E, as I am a bit worried if the is no confusion between $U$ and $U^\dagger$}


\begin{thebibliography}{99}
\bibitem{PAN}
J.-W. Pan, Z. B. Chen, C. Y. Lu, H. Weinfurter, A. Zeilinger, and M. \.{Z}ukowski, Rev. Mod. Phys. {\bf 84}, 777 (2012).

\bibitem{ASPECT}
A. Aspect, P. Grangier, and G. Roger, Phys. Rev. Lett. {\bf 49}, 91 (1982).

\bibitem{GHZ}
D. M. Greenberger, M. A. Horne, and A. Zeilinger,
Bell's Theorem, Quantum Theory, and Conceptions of the Universe, M. Kafatos (Ed.), 
Kluwer, Dordrecht (1989), 69-72.

\bibitem{BANASZEK}
K. Banaszek and K. W{\'o}dkiewicz, Phys. Rev. A {\bf 58}, 4345 (1998);
%https://doi.org/10.1103/PhysRevA.58.4345
Acta. Phys. Slov. {\bf 49}, 491-500 (1999).

\bibitem{BOUW}
C. Simon and D. Bouwmeester, Phys. Rev. Lett. {\bf 91}, 053601 (2003).

\bibitem{MASZA-REVIEW}
M. V. Chekhova, G. Leuchs, and M. \.{Z}ukowski, Opt. Commun. {\bf 337}, 27 (2015).

\bibitem{EPR}
A. Einstein, B. Podolsky, and N. Rosen, Phys. Rev. {\bf 47}, 777 (1935).

\bibitem{ROSOLEK}
K. Roso\l ek, K. Kostrzewa, A. Dutta, W. Laskowski, M. Wie\'sniak, and M. \.Zukowski,
Phys. Rev. A {\bf 95}, 042119 (2017).

\bibitem{DURKIN}
G. A. Durkin, C. Simon, and D, Bouwmeester,
Phys. Rev. Lett. {\bf 88}, 187902 (2002).

\bibitem{HORODECKIS}
R. Horodecki, P. Horodecki, M. Horodecki, and K. Horodecki, Rev. Mod. Phys. {\bf 81}, 865 (2009).

\bibitem{SIMON}
R. Simon, Phys. Rev. Lett. {\bf 84}, 2726 (2000).

\bibitem{DUAN}
L. -M Duan, et al., Phys. Rev. Lett. {\bf 84}, 2722 (2000).

\bibitem{HILLERY}
M. Hillery and M. S. Zubairy, Phys. Rev. Lett. {\bf 96}, 050503 (2006).

\bibitem{supp}
See Supplemental Material at [URL] for more details.

\bibitem{Iskhakov12}
T. Sh. Iskhakov, I. N. Agafonov, M. V. Chekhova and G. Leuchs, Phys. Rev. Lett. {\bf 109}, 150502 (2012).

\bibitem{Kanseri13}
B. Kanseri, T. Iskhakov, G. Rytikov, M. Chekhova and G. Leuchs, Phys. Rev. A {\bf 87}, 032110 (2013).

\bibitem{BOUW-2}
H. S. Eisenberg, G. Khoury, G. A. Durkin, C. Simon, D. Bouwmeester, Phys. Rev. Lett. {\bf 93}, 193901 (2004).

\bibitem{Iskhakov13}
T. Sh. Iskhakov, K. Y. Spasibko, M. V. Chekhova and G. Leuchs, New Journal of Physics {\bf 15}, 093036 (2013).

\bibitem{Lamas01}
A. Lamas-Linares, J. C. Howell, D. Bouwmeester, Nature {\bf 412}, 887–890 (2001).

%\bibitem{Eisenberg04}
%H. S. Eisenberg, G. Khoury, G. A. Durkin, C. Simon, and D. Bouwmeester, Phys. Rev. Lett. {\bf 93}, 193901 (2004).

\bibitem{Eckstein11}
A. Eckstein, A. Christ, P. J. Mosley and C. Silberhorn, Phys. Rev. Lett. {\bf 106}, 013603 (2011).

\bibitem{Spasibko17}
K. Yu. Spasibko, D. A. Kopylov, V. L. Krutyanskiy, T. V. Murzina, G. Leuchs and M. V. Chekhova, Phys. Rev. Lett. {\bf 119}, 223603 (2017).

\bibitem{Mattle95}
K. Mattle, M. Michler, H. Weinfurter, A. Zeilinger and M. \.{Z}ukowski, Appl. Phys. B {\bf 60}, S111–7 (1995).

\bibitem{Weihs96}
G. Weihs, M. Reck, H. Weinfurter and A. Zeilinger, Phys. Rev. A {\bf 54}, 893 (1996).

\bibitem{Peruzzo11}
A. Peruzzo, A. Laing, A. Politi, T. Rudolph and J. L. O’Brien, Nat. Commun. {\bf 2}, 224 (2011).

\bibitem{Meany12}
T. Meany, M. Delanty, S. Gross, G. D. Marshall, M. J. Steel and M. J. Withford, Opt. Express {\bf 20}, 26895 (2012).

\bibitem{Metcalf13}
B. J. Metcalf et al. Nat. Commun. {\bf 4}, 1356 (2013).

\bibitem{Spagnolo13}
N. Spagnolo, C. Vitelli, L. Aparo, P. Mataloni, F. Sciarrino, A. Crespi, R. Ramponi and R. Osellame, Nat. Commun. {\bf 4}, 1606 (2013).

\bibitem{Carolan15}
J. Carolan et al., Science {\bf 14}, 711 (2015).

\bibitem{Schaeff15}
C. Schaeff, R. Polster, M. Huber, S. Ramelow and A. Zeilinger, Optica {\bf 2}, 523 (2015).

\bibitem{SORENSEN} 
A. S{\o}rensen, L.-M. Duan, J. I. Cirac and P. Zoller, Nature {\bf 409}, 63 (2001).

\bibitem{POLZIK} 
J. Hald, J. L. S{\o}rensen, C. Schori, and E. S. Polzik, Phys. Rev. Lett. {\bf 83}, 1319 (1999).

\bibitem{sorensenPRL}
A. S. S{\o}rensen and K. M{\o}lmer, Phys. Rev. Lett. {\bf 86}, 4431 (2001).

\bibitem{ZUK-ZEIL-WEIN}
M. {\.Z}ukowski, A. Zeilinger and H. Weinfurter, Ann. NY Acad. Sci. {\bf 755}, 91 (1995).

\bibitem{KALTENBAEK}
R. Kaltenbaek, B. Blauensteiner, M. {\.Z}ukowski, M. Aspelmeyer, and A. Zeilinger, Phys. Rev. Lett. {\bf 96}, 240502 (2006).

\bibitem{zuko}
M. {\.Z}ukowski, W. Laskowski, and M. Wie{\'s}niak, Phys. Rev. A {\bf 95}, 042113 (2017).

\bibitem{zuko1}
M. {\.Z}ukowski, W. Laskowski, and M. Wie{\'s}niak, Phys. Scr. {\bf 91}, 084001 (2016).

\bibitem{zuko2}
M. {\.Z}ukowski, M. Wie{\'s}niak, and W. Laskowski, Phys. Rev. A {\bf 94}, 020102(R) (2016).

\bibitem{HE}
Q. Y. He, M. D. Reid, T. G. Vaughan, C. Gross, M. Oberthaler, and P. D. Drummond, Phys. Rev. Lett.
{\bf 106}, 120405 (2011); Q. Y. He, T. G. Vaughan, P. D. Drummond, and M. D. Reid, New Journal of Physics {\bf 14}, 093012 (2012).

%\bibitem{footnote1}
%Operationally, in the $r$-th run of an experiment, this requires measurement of photon numbers in the two exits of a polarization analyzer, $n^r_j$ and $n^r_{j_\perp}$, and dividing their difference by their sum. If the sum is zero, the value is put as zero. The procedure does not require any additional measurements, only the data are differently processed compared to the standard approach. In~\cite{zuko, zuko1, zuko2}, some examples of two-party entanglement conditions and Bell inequalities using such normalized Stokes operators were given. Here we  present a different generalized approach, which allows direct extensions to more involved cases.

%\bibitem{footnote2}
%However, as a purity of a field state $\ket{\psi^{AB}_{\lambda}}$ does not warrant that the corresponding $\hat{R}^{AB}_{\lambda}$ is a projector, $ \mathbf{\mathfrak{\hat R}}^{AB}_\varrho$ does not have to have the same convex expansion coefficients in terms of pure two-qubit states, as $\varrho$ in terms of $\ket{\psi^{AB}_{\lambda}}$'s.

%\bibitem{footnote3}
%This employs a perfect detector in front of which is a beamsplitter of transmitivity amplitude ${\eta}$, with the reflection channel describing the losses.

\bibitem{RECK}
M. Reck, A. Zeilinger, H. J. Bernstein, and P. Bertani, Phys. Rev. Lett. {\bf 73}, 58 (1994).

\bibitem{WOOTERS}
W. K. Wootters and B. D. Fields, Ann. Phys. {\bf 191}, 363 (1989).

\bibitem{MUBs}
I. D. Ivanovic, J. Phys. A: Math. Theor. {\bf 14}, 3241 (1981).



\bibitem{BELL-RMP}
N. Brunner, D. Cavalcanti, S. Pironio, V. Scarani, and S. Wehner,
Rev. Mod. Phys. {\bf 86}, 419 (2014).

%\bibitem{CHINY}
%S. Yu, J.-W. Pan, Z.-B. Chen, and Y.-D. Zhang, Phys. Rev. Lett. {\bf 91}, 217903 (2003).

%\bibitem{zukoarxiv}
%M. {\.Z}ukowski, W. Laskowski, and M. Wie{\'s}niak, arXiv:1508.02368v2

%\bibitem{SIMON} 
%H. S. Eisenberg, G. Khoury, G. A. Durkin, C. Simon, and D. Bouwmeester, Phys. Rev. Lett. {\bf 93}, 193901 (2004).

%\bibitem{KLYSHKO}
%D. M. Klyshko, J. Exp. Theor. Phys. {\bf 84}. 1065 (1997).
% https://doi.org/10.1134/1.558243

%\bibitem{MULTIPORT}
%J. Ryu, M. Marciniak, M. Wie\'{s}niak, and M. {\.Z}ukowski, J. Opt. {\bf 20}, 044002 (2018).

%\bibitem{MULTIPORT2}
%J. Ryu, M. Marciniak, M. Wie\'{s}niak, D. Kaszlikowski, and M. {\.Z}ukowski, Acta Phys. Pol. A {\bf 132}, 1713 (2017).

\bibitem{SCHMIED}
R. Schmied, J.-D. Bancal, B. Allard, M. Fadel, V. Scarani, P. Treutlein, and N. Sangouard,
Science {\bf 352}, 441 (2016).
%DOI: 10.1126/science.aad8665

%%%%%%%%%%%%%%%
% Refs. for appendix
%%%%%%%%%%%%%%%

%\bibitem{SIMON} 
%H. S. Eisenberg, G. Khoury, G. A. Durkin, C. Simon, and D. Bouwmeester, Phys. Rev. Lett. {\bf 93}, 193901 (2004).

%\bibitem{BOUW}
%C. Simon and D. Bouwmeester, Phys. Rev. Lett. {\bf 91}, 053601 (2003).



%\bibitem{BOUW-2} H. S. Eisenberg, G. Khoury, G. A. Durkin, C. Simon, D. Bouwmeester,
%Phys. Rev. Lett. {\bf 93}, 193901 (2004)

%\bibitem{zuko}
%M. {\.Z}ukowski, W. Laskowski, and M. Wie{\'s}niak, Phys. Rev. A {\bf 95}, 042113 (2017).

\bibitem{MULTIPORT}
J. Ryu, M. Marciniak, M. Wie\'{s}niak, and M. {\.Z}ukowski, J. Opt. {\bf 20}, 044002 (2018).

\bibitem{MULTIPORT2}
J. Ryu, M. Marciniak, M. Wie\'{s}niak, D. Kaszlikowski, and M. {\.Z}ukowski, Acta Phys. Pol. A {\bf 132}, 1713 (2017).

\bibitem{CHINY}
S. Yu, J.-W. Pan, Z.-B. Chen, and Y.-D. Zhang, Phys. Rev. Lett. {\bf 91}, 217903 (2003).

%\bibitem{zuko1}
%M. {\.Z}ukowski, W. Laskowski, and M. Wie{\'s}niak, Phys. Scr. {\bf 91}, 084001 (2016).

%\bibitem{Iskhakov12}
%T. Sh. Iskhakov, I. N. Agafonov, M. V. Chekhova and G. Leuchs, Phys. Rev. Lett. {\bf 109}, 150502 (2012).

\bibitem{KLYSHKO}
J.  M. Jauch and F. Rohrlich, {\it The Theory of Photons and Electrons} (Addison-Wesley, Reading, MA, 1955).

\end{thebibliography}
\end{document}